\documentclass[twocolumn,aps,amsmath,pre,floatfix,superscriptaddress]{revtex4-2}
\usepackage{graphicx}
\usepackage{comment}
\usepackage{geometry}
\usepackage{wrapfig}
\usepackage{comment}
\usepackage{amsmath}
\usepackage[export]{adjustbox}
\usepackage{amssymb}
\usepackage[utf8]{inputenc}
\usepackage{afterpage}
\usepackage{lipsum}  
\usepackage{appendix}

\usepackage{hyperref}
\hypersetup{colorlinks=true, linktoc=all, linkcolor=blue, linktocpage, citecolor=blue}

\usepackage{geometry}\usepackage{geometry}

\usepackage[normalem]{ulem}
\usepackage{xcolor}

\graphicspath{{figs/}}

\begin{document}
\title{Heterogeneous contributions can jeopardize cooperation in the Public Goods Game }


\author{Lucas S. Flores}
\author{Mendeli H. Vainstein}
\author{Heitor C. M. Fernandes}
\address{Instituto de Física, Universidade Federal do Rio Grande do Sul, CP 15051, CEP 91501-970 Porto Alegre - RS, Brazil}
\author{Marco A. Amaral}
\address{Instituto de Humanidades, Artes e Ciências, Universidade Federal do Sul da Bahia, CEP 45638-000, Teixeira de Freitas - BA, Brazil.}

\date{\today}

\begin{abstract}
When studying social dilemma games, a crucial question arises regarding the impact of general heterogeneity on cooperation, which has been shown to have positive effects in numerous studies.
Here, we demonstrate that heterogeneity in the contribution value for the focal Public Goods Game can jeopardize cooperation.     
We show that there is an optimal contribution value in the homogeneous case that most benefits cooperation depending on the lattice.
In a heterogeneous scenario, where strategy and contribution coevolve, cooperators making contributions higher than the optimal value end up harming those who contribute lower.  
 This effect is notably detrimental to cooperation in the square lattice with von Neumann neighborhood, while it can have no impact in others lattices. 
Furthermore, in parameter regions where a higher-contributing cooperator cannot normally survive alone, the exploitation of lower value contribution cooperators allows their survival, resembling a parasitic behavior.
To obtain these results, we employed various distributions for the contribution values in the initial condition and conducted Monte Carlo simulations.        

\end{abstract}


\maketitle

\section{Introduction}
\label{intro}

Evolutionary Game Theory provides a mathematical and theoretical framework to understand the dynamics of cooperative behavior in a competitive environment. In this framework, individuals are seen as rational players who usually interact with each other aiming to maximize their own gains. One of the central questions in this field is how cooperation can emerge and persist in such a context, where individuals are motivated by self-interest and are exposed to the pressures of natural selection~\cite{bergstrom2002evolution}. 
Despite the numerous contributions made to this field, the underlying mechanisms  which allow cooperation continue to be the subject of intensive research.

Public Goods Games (PGGs) are a classic model used in game theory to study the evolution of cooperation. In a PGG, participants must decide how much to contribute to a common pool of resources. The total contribution is multiplied by a factor greater than one, reflecting the positive feedback of cooperation, and then divided equally among all participants, regardless of their individual contribution. This creates a dilemma, as participants have an incentive to free-ride and not contribute, while the overall outcome is improved by cooperation.
An  example of a real-world interaction that is often described by the dynamics of PGGs is the provision of public goods and services, such as education, health care, and environmental protection~\cite{gois2019reward}. In these contexts, individuals must decide whether to contribute to the provision of these goods and services, and their decision is influenced by the level of contributions made by others.

Many mechanisms have been proposed to sustain cooperation, such as memory~\cite{santos2011evolution}, spatial reciprocity~\cite{ohtsuki2006simple,tarnita2009strategy,Wang2012,hindersin2014counterintuitive,allen2014games, FLORES2022112744}, punishment~\cite{FLORES2021110737,brandt2003punishment,chen2014probabilistic}, reward~\cite{szolnoki2010reward}, aspiration~\cite{amaral2016stochastic, shi2021dynamic}, commitment to contribute~\cite{han2017evolution}, voluntary interactions~\cite{hauert2003prisoner}, and heterogeneity~\cite{perc2015double, perc2011does, shi2010group,attila_heterog, attila_heterog2, marco_heterog, CAO20101273, kun2013resource}.

Typically, studies of the PGG assume that all players contribute equally. However, this assumption does not reflect real-world situations where the distribution of wealth in society is heterogeneous, with some individuals possessing significantly more resources than others.
To address this issue, researchers have explored the impact of heterogeneity in the distribution of contribution values on cooperation levels in the PGG. For instance, some studies have examined situations where the contribution depends on the number of cooperators ($C$s) in the group~\cite{wang2018heterogeneous, yuan2014role}, and have observed an increase in cooperation levels.
In addition, was investigated the effect of using a uniform distribution for the contribution of cooperators~\cite{huang2015effect} and  demonstrated that increasing the range of the distribution can lead to a greater benefit for cooperation.
Heterogeneity in players' contribution, depending on the players' age, was explored in~\cite{TIAN201365} yielding an enhancement in cooperation for some situations.

At a first glance, a topic that may appear unrelated to heterogeneity is the effect of noise in regular lattices~\cite{Szolnoki_2009,Szab__2005, Vukov_2006}. However, as we will demonstrate in this paper, different contribution values in the PGG can be interpreted as distinct noise scenarios when the update rule follows the Fermi function. 
In this work, we investigate the PGG in the classical homogeneous scenario where all cooperators contribute a fixed  value of $c$ and  explore how it affects the amount of cooperation. When dealing with heterogeneous scenarios, we assume that initially each cooperator contributes according to a given distribution, while defectors consistently contribute nothing. A crucial aspect of our model is that the players always copy the contribution value when switching strategies.
Through Monte Carlo simulations,  we reveal that contribution heterogeneity can actually hinder cooperation in some parameter regions. However, our study also uncovers interesting phase transitions and second-order free-riding effects.  These findings shed new light on the complex relationships between wealth disparity, cooperation, and the provision of public goods and services.


\section{Model} 
\label{model}

Here, we study the heterogeneous Focal Public Goods Game (FPGG, also called pairwise interactions~\cite{perc2013evolutionary}), in which cooperators ($C$) may contribute distinct values,  whereas defectors ($D$) contribute nothing. For the FPGG in a regular lattice, a player's payoff, $\Pi_X$,  is calculated by summing the contributions from all their first neighbors, including themselves. 
Next, all contributions are multiplied by a factor $r$ and the resulting value  is equally distributed among all $G$ members of the group. Finally, each cooperator pays a cost,  thus resulting in 
\begin{align}
    \Pi_{C_{c_i}} &= \frac{r}{G} \sum_{k=1}^{G} c_k - c_i \label{eqC}\\
    \Pi_{D} &= \frac{r}{G} \sum_{k=1}^{G} c_k  \label{eqD},
\end{align}
where $c_k$ is player $k$'s contribution  ($c_k=0$ if it is a defector) and we denote the central player and their contribution by $C_{c_i}$ and $c_i$, respectively.
We initialize the simulation with an equal fraction of randomly distributed cooperators and defectors. Uniform,  Gaussian,  and Bernoulli discrete distributions of the contributions are used among the cooperating half of the initial population when dealing with the heterogeneous scenario. When referring to a homogeneous case,  we use a fixed $c_i \equiv c$ for all cooperators in the population.

The system evolves as follows: first, a player $X$ and one of their neighbors $Y$ are randomly chosen, and their payoffs are calculated from the equations above. Player $X$ adopts $Y$'s strategy, and contribution value ($c_Y$) with probability 
\begin{equation} \label{eq.transition}
     W_{X\rightarrow Y}= \frac{1}{1+e^{-(\Pi_Y-\Pi_X)/K}} \,\,,   
\end{equation}
where $K$ is the noise associated with irrationality. The same process is repeated $N$ times characterizing one Monte Carlo Step (MCS), where $N$ is the total population size. We used $t_{max} = 10^5$ MCS and $K=0.1$. Averages were performed over $100$ random initial conditions where appropriate.
The simulations were performed on the square lattice with von Neumann neighborhood ($G=5$) and on the triangular lattice ($G=7$), both with $N=100^2$ and periodic boundary conditions. 


\section{Results} 
\label{results}

\subsection{Homogeneous case}

\begin{figure*}[t]
    \centering
    \includegraphics[width=0.75\columnwidth]{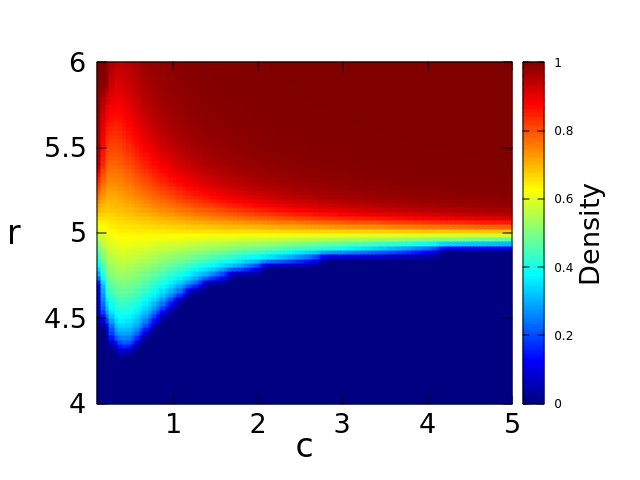}
    \includegraphics[width=0.75\columnwidth]{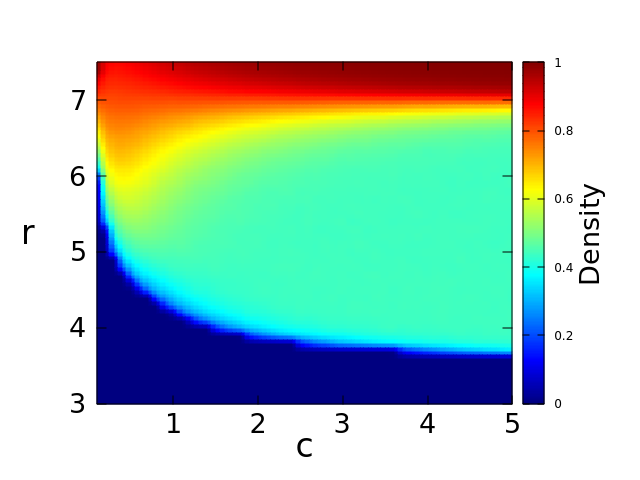}
    \caption{
     The phase diagrams of the equilibrium cooperation density for the square lattice with von Neumann neighborhood (left panel)  and for the triangular lattice (right panel) show the combined  effects of the multiplicative factor, $r$, and the contribution value, $c$, on cooperation for homogeneous populations. These diagrams illustrate that there exists an optimal contribution value ($c_o \approx 0.4$)  that minimizes the corresponding critical $r$ value above which cooperation can occur for the square lattice case ($r_{c_o} \approx 4.3$). On the other hand, for the triangular lattice, the critical $r$ value decreases monotonically as the contribution increases.
    }
    \label{diagrama}
    
\end{figure*}

First, we address the effect of a homogeneous contribution value for all cooperators in the population.  In this case, Eqs.~(\ref{eqC}) and (\ref{eqD}) become
\begin{align}
    \Pi_{C} &= \frac{rc}{G} \, N_C^C  - c \label{eq.homoc}\\ 
    \Pi_{D} &= \frac{rc}{G} \, N_C^D, \label{eq.homod}
\end{align}
where  $N_C^i$ are the number of cooperators in player $ i$'s  group and contribute the same value $c$ each.  If we substitute  Eqs.~(\ref{eq.homoc}) and (\ref{eq.homod}) into  Eq.~(\ref{eq.transition}),  $c$ factors out enabling us to define an effective noise $K^\prime = K/c$.
Therefore, a simulation with fixed $K=0.1$ but with a varying contribution value $c$ can be interpreted as viewing the system under different noise scenarios. 
Fig.~\ref{diagrama}  shows the phase diagram ($r \times c$) for the steady-state densities of cooperation on the square (left panel) and triangular (right panel) lattices. 
For each $c$, there is a corresponding critical value $r_c$ below which cooperation is unviable.
For the square lattice, there is an optimal intermediate contribution value ($c_o\approx 0.4$,  i.e, $K^\prime \approx 0.25$), that sustains cooperation with  $r_{c_o} \approx 4.3$. 
 In contrast, the triangular lattice exhibits a distinct behavior, where $r_c$ decreases monotonically as the contribution value $c$ increases.
Both observations are consistent with  well-known results: 
previous studies have established that regular lattices with zero clustering coefficient, such as the square lattice, exhibit an optimal noise value that can sustain cooperation~\cite{Szab__2005,Vukov_2006,perc2013evolutionary}. Conversely, in lattices with non-null clustering coefficients, such as the triangular lattice, a  deterministic scenario  is the best for cooperation to persist, in the sense that the critical  $r$ tends to the lowest values in this case. 

 A qualitative understanding of the above results can be achieved by recalling that the square lattice has only one important deterministic transition, occurring at $r=G$, whereas the triangular lattice has two such transitions at $r=G/2$ and $G$, as illustrated in the appendix of~\cite{FLORES2022112744}. 
In the regime where $r>G$, cooperation dominates in the low-noise case ($c\gg1$) for both lattices. Thus, the introduction of noise can either harm cooperation or maintain the domination unchanged. On the other hand, for low values of $r$ such that $r<G$ for the square lattice and $r<G/2$ for the triangular lattice, cooperation becomes extinct, and therefore the introduction of noise can only benefit cooperators or maintain their extinction.

For the triangular lattice, between these limits ($G/2<r<G$), cooperators coexist with defectors in the low noise case.  
Here, the introduction of noise can have different effects, depending on the parameter values. This is, in fact, the case since cooperation is favored slightly below $r=G$  and compromised above  the $r=G/2$ transition. 
Were if not for this second transition at $r=G/2$, the results for the triangular lattice would be similar to those of the square lattice since it also has a locally optimal intermediate contribution value if we restrain ourselves to the range $r\gtrapprox 5.5$ and the phase diagrams are similar above this value, presenting a maximum in cooperation density in both cases if $r<G$.

\begin{figure*}[t]
\begin{center}
\includegraphics[width=.36\columnwidth]{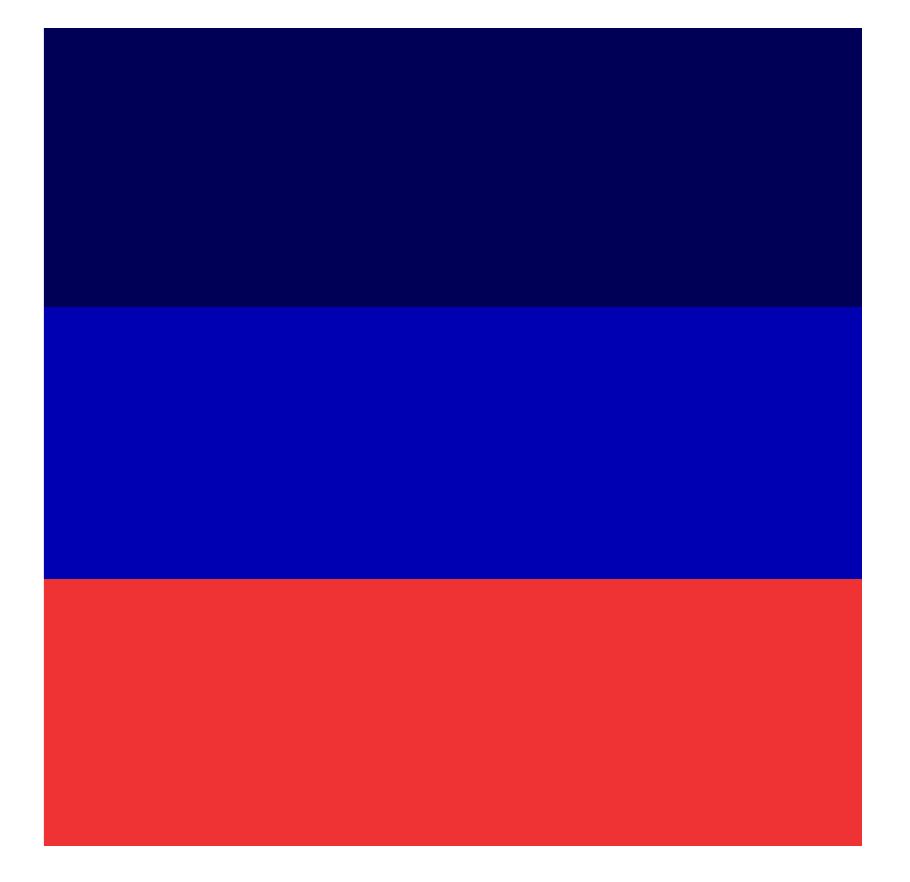}
\includegraphics[width=.36\columnwidth]{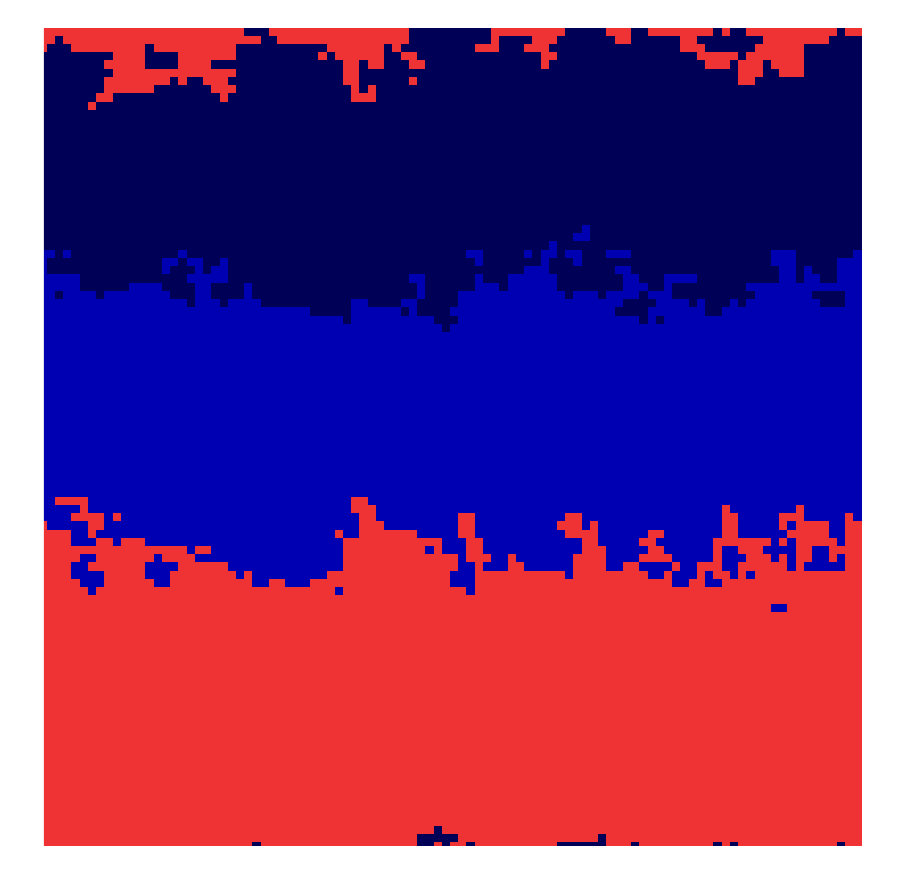}
\includegraphics[width=.36\columnwidth]{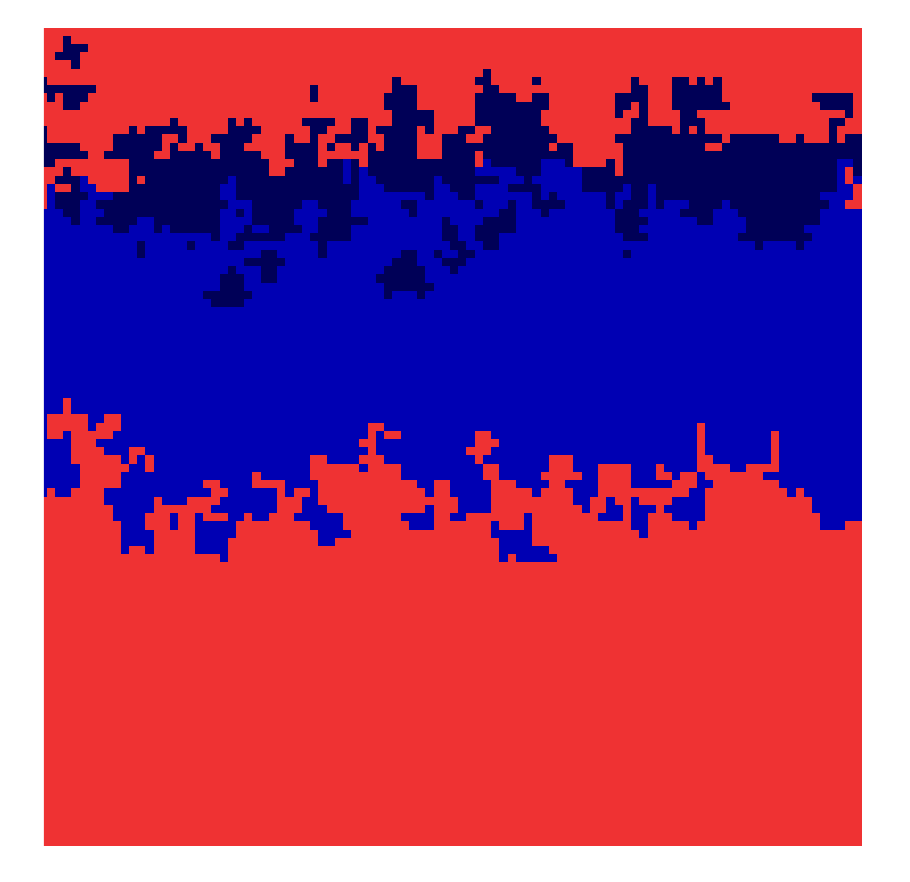}
\includegraphics[width=.36\columnwidth]{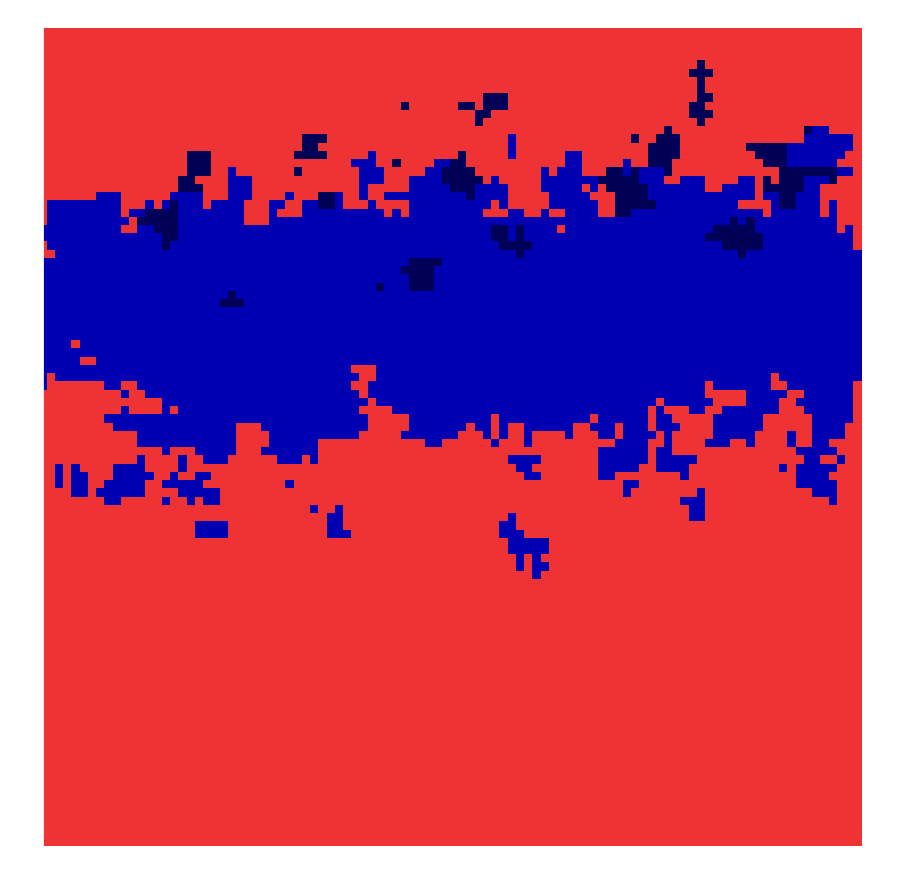}
\includegraphics[width=.36\columnwidth]{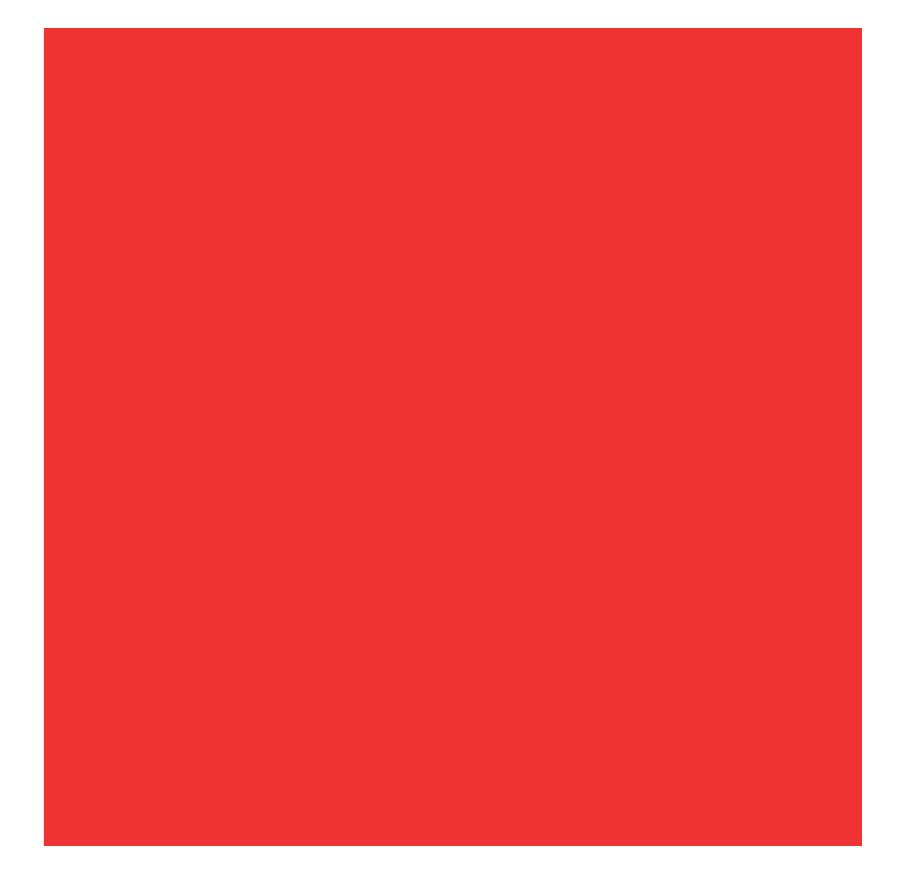}

\includegraphics[width=.36\columnwidth]{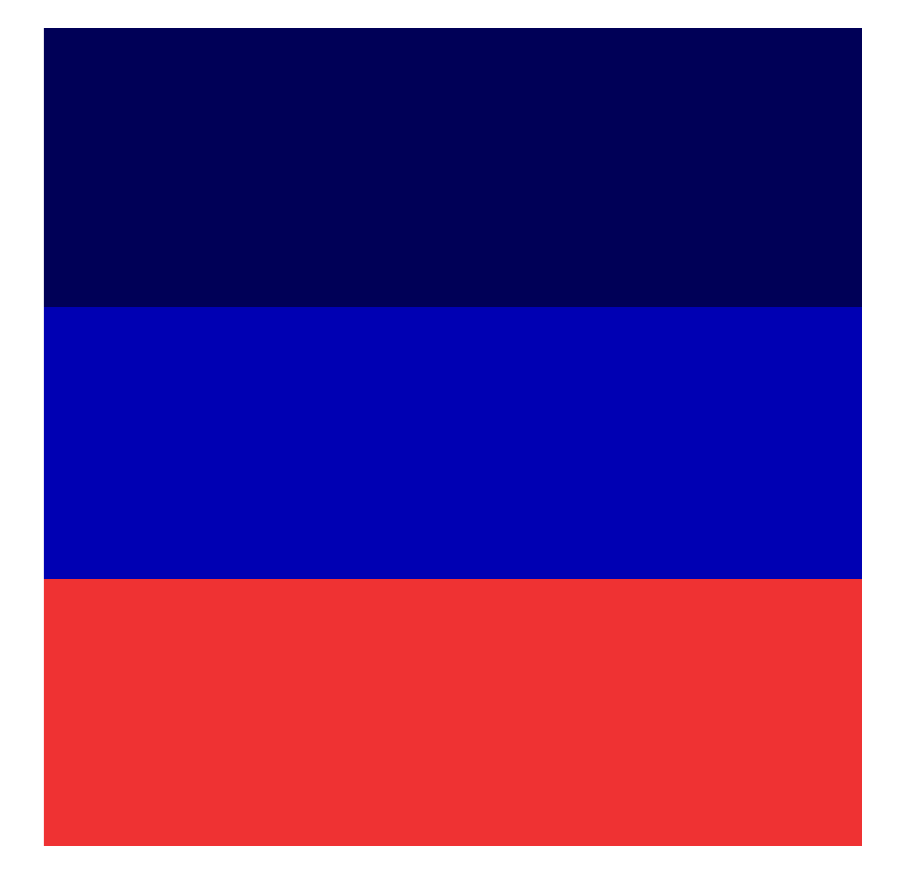}
\includegraphics[width=.36\columnwidth]{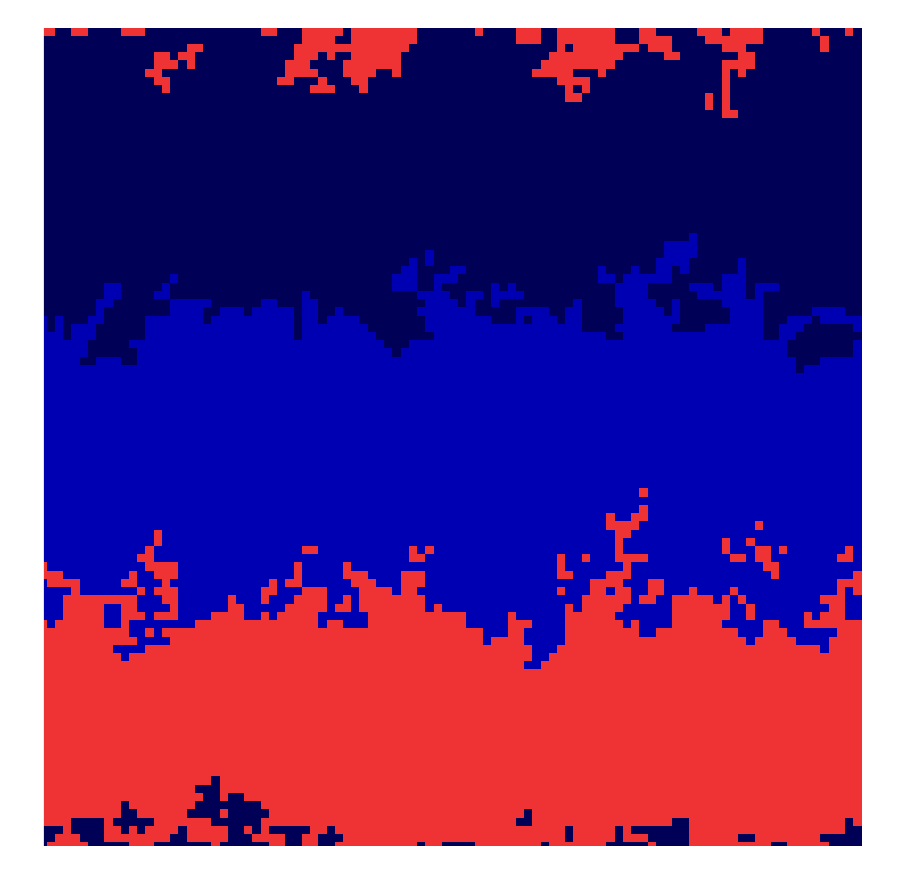}
\includegraphics[width=.36\columnwidth]{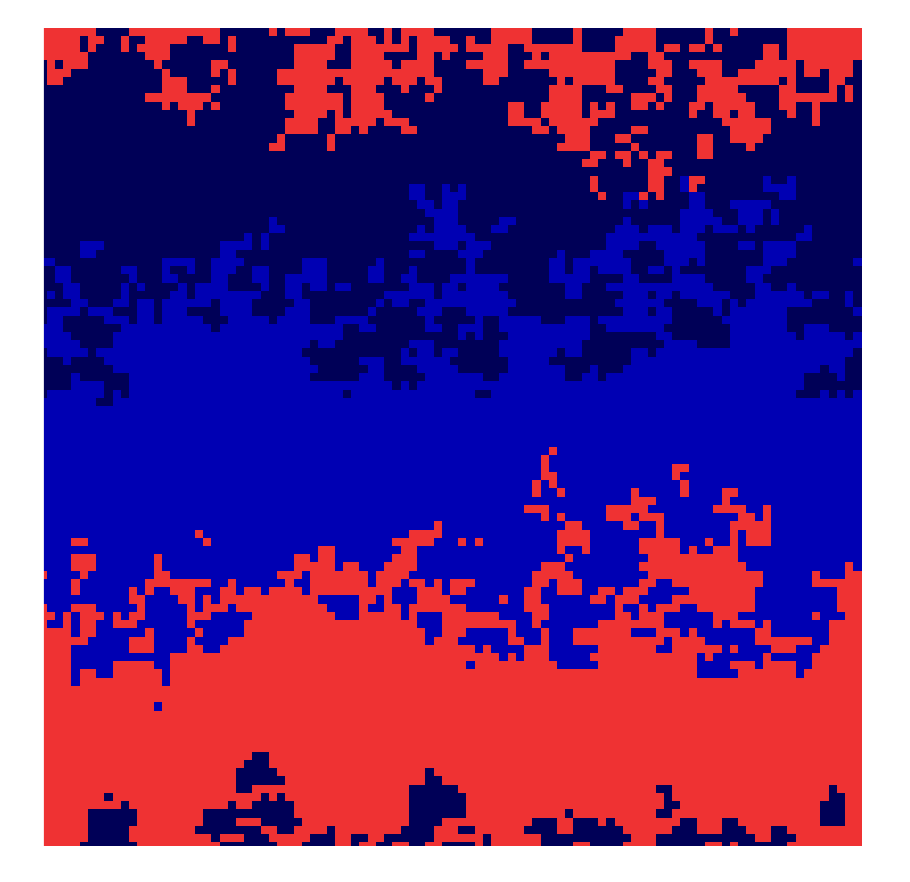}
\includegraphics[width=.36\columnwidth]{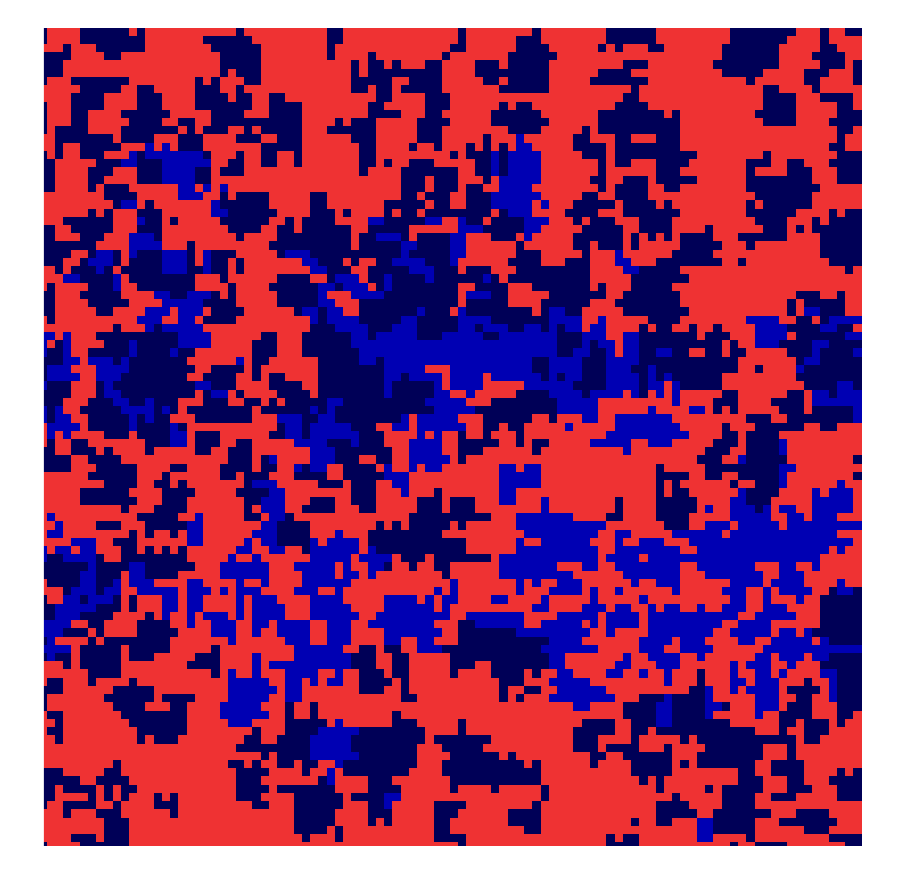}
\includegraphics[width=.36\columnwidth]{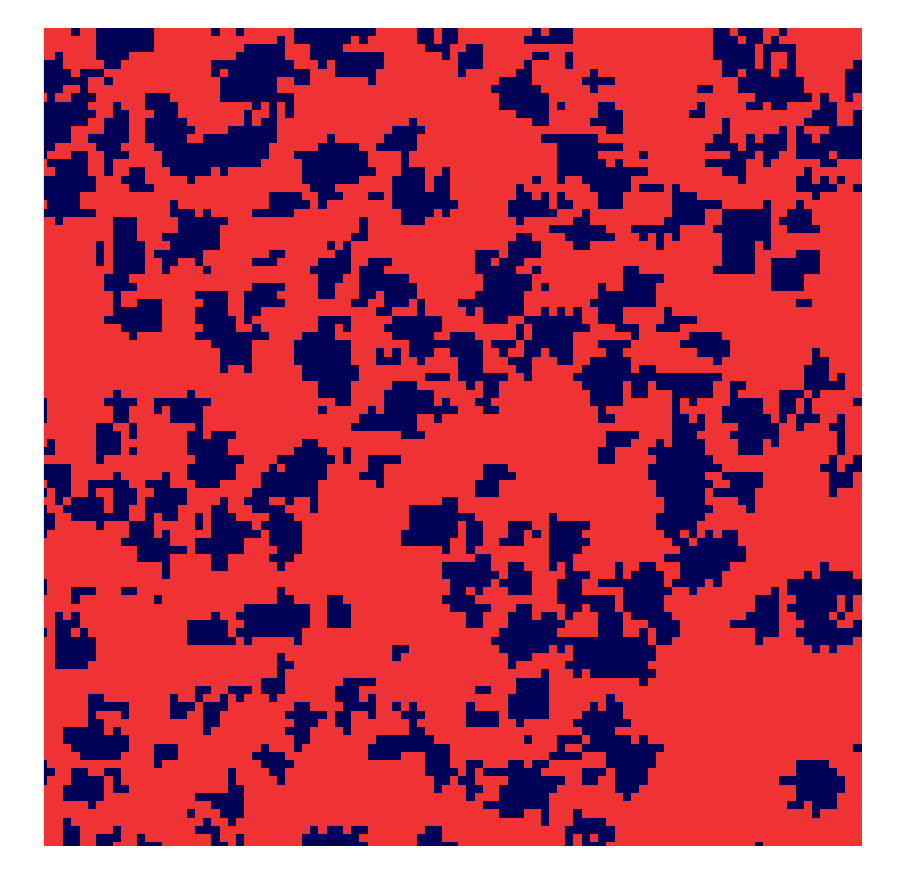}

\includegraphics[width=.36\columnwidth]{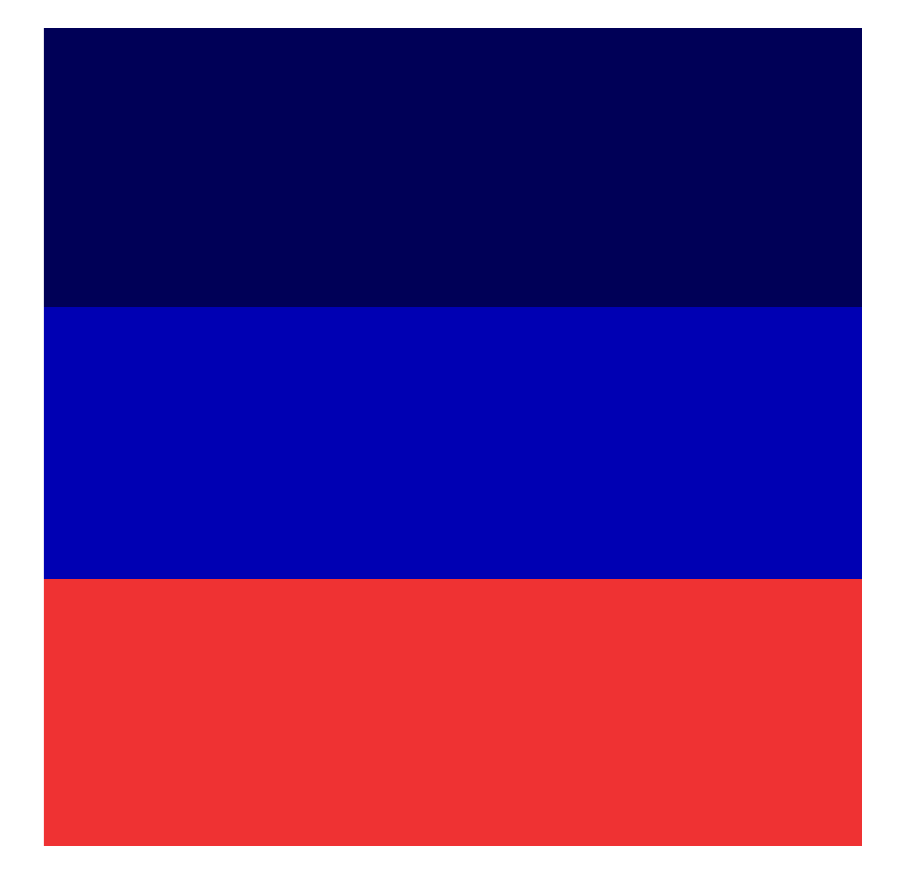}
\includegraphics[width=.36\columnwidth]{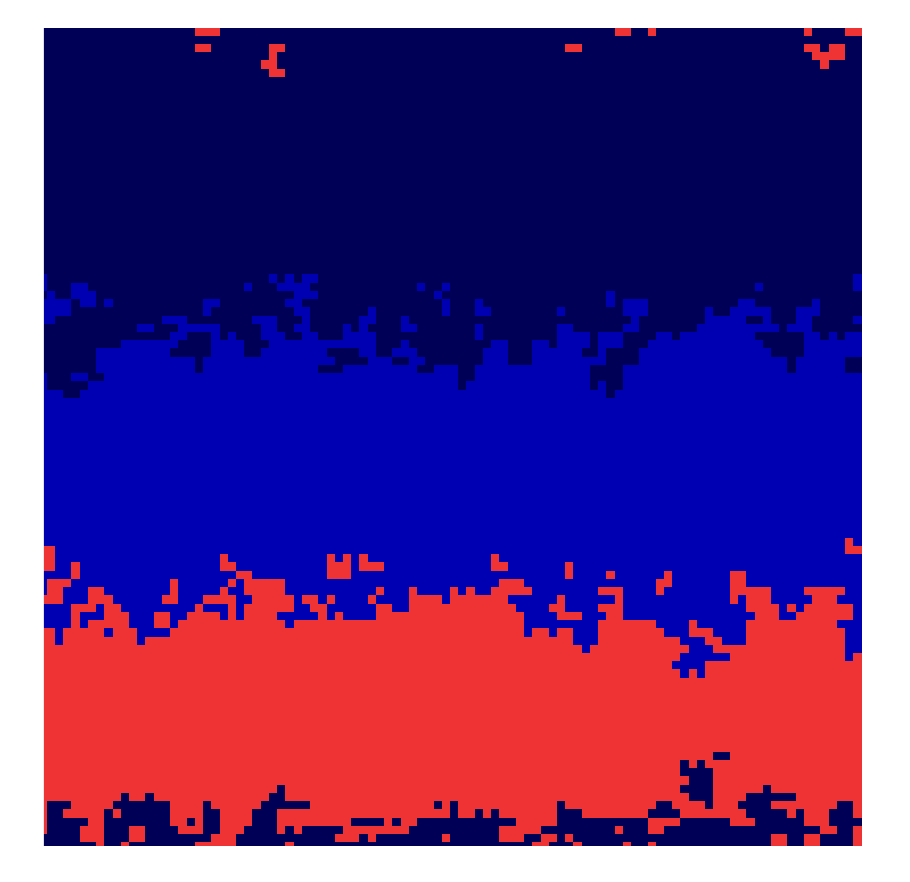}
\includegraphics[width=.36\columnwidth]{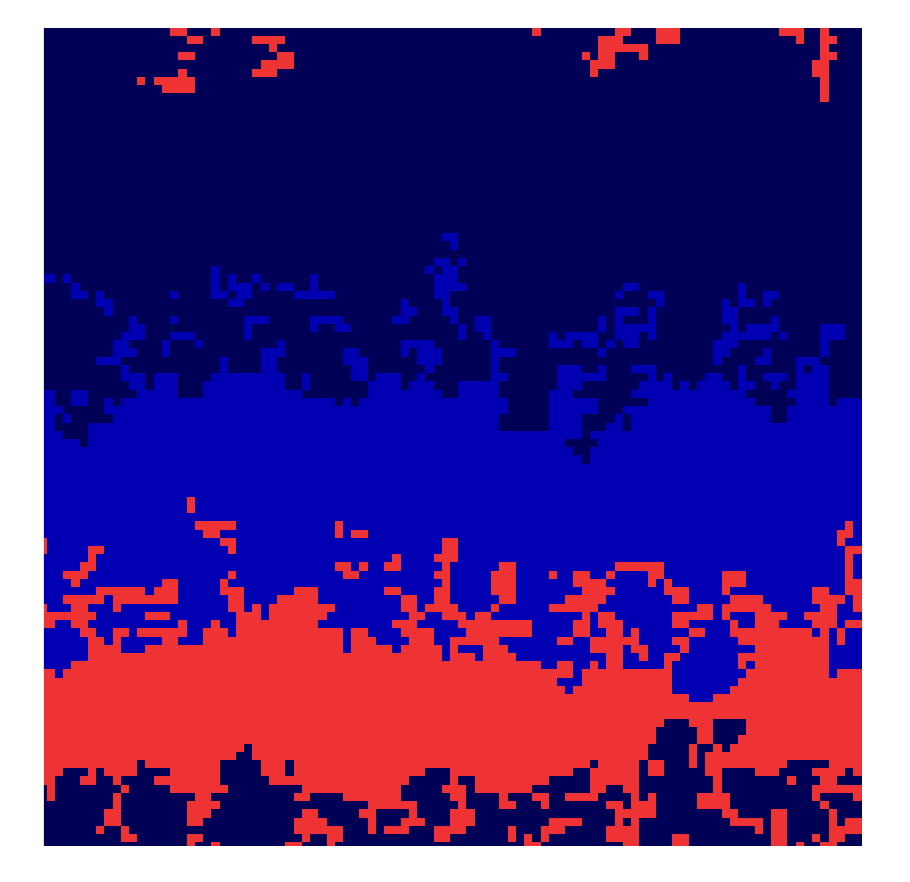}
\includegraphics[width=.36\columnwidth]{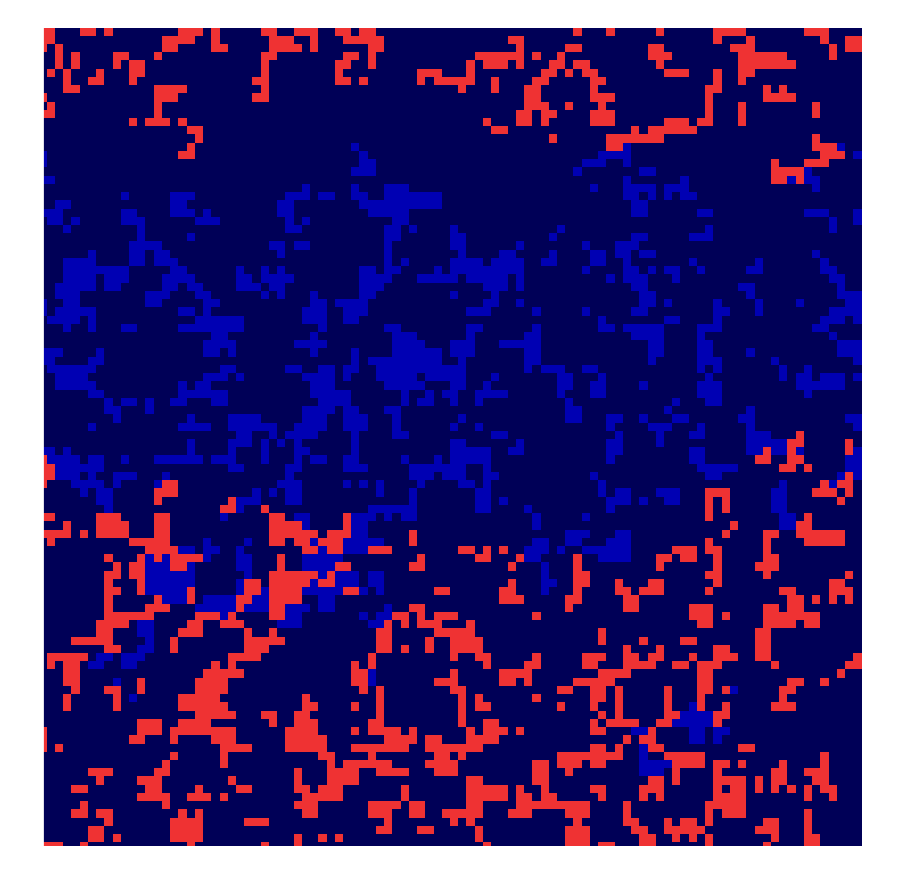}
\includegraphics[width=.36\columnwidth]{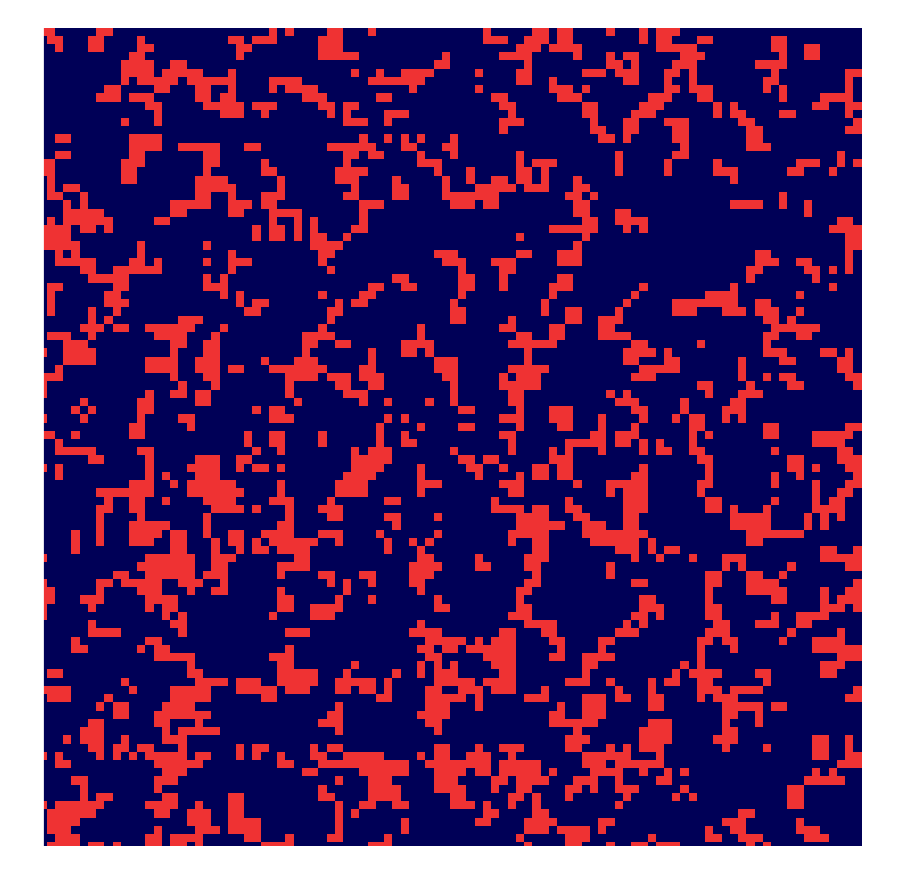}

    \caption{
    Temporal evolution of cooperators in clusters with different $c$ values on the square lattice: $c=0.5$ (light red), $c=1.0$ (light blue), and $c=2$ (dark blue). The top row shows the dominance of the lower contribution cooperators for $r=3.7$. The middle row illustrates the coexistence between the lowest and the highest contribution cooperators for $r=4.78$. Finally, the bottom row illustrates the regime  where higher contributing cooperators begin to dominate ($r=5.1$).  
   The number of MCS steps taken during each time evolution vary, highlighting the unique characteristics of each situation. The last column corresponds to $300$ MCS for the first row and $1300$ MCS for the other two. 
    }
    \label{corte}
    \end{center}
\end{figure*}

\subsection{Heterogeneous case}
\label{het}

Consider a population consisting solely of cooperators with varying contribution values. In this scenario, when a player adopts a neighbor's strategy, they also adopt their contribution value. As a result, cooperators with similar $c_i$ values tend to cluster together. We can use the phase diagrams presented in Fig.~\ref{diagrama} to gain insight into the behavior of this heterogeneous population.

Fig.~\ref{corte} illustrates a specially prepared initial condition of cooperators, $C_{c_i}$,  with different contribution values ($c_i$): $0.5$ (light red), $1$ (light blue), and $2$ (dark blue), in the absence of defectors. In general, we observe that lower contribution cooperators behave similarly to defectors. Consequently, for sufficiently small $r$, the $C_{0.5}$ cooperators dominate the population (top row). 
When $r$ is sufficiently large, we observe coexistence between $C_{0.5}$ and $C_{2}$ (middle row). In this case, $C_{2}$ forms clusters within a sea of $C_{0.5}$, creating a situation that resembles the classic $C$ \textit{vs} $D$ game. For smaller $r$ values, where $C_2$ cannot survive, we find that $C_{0.5}$ and $C_{1}$ can also coexist.
Moreover, the second row in Fig.~\ref{corte} demonstrates that $C_1$ can initially coexist with both $C_{0.5}$ and $C_2$ individually, but its density eventually declines until it becomes extinct when all three contribution values are present. This implies that $C_{0.5}$ and $C_2$ exhibit some synergistic effect that jeopardizes the survival of $C_1$ cooperators. This observation is crucial and will be further explored in the subsequent sections of this paper.
Finally, for larger $r$ values, we see that the $C_2$ starts to predominate in the population (bottom row).
An important observation is that the higher contribution cooperators always cluster within the sea of lower contribution cooperators, suggesting that the notion of cooperators and defectors depends on whom the game is played  against. In Appendix~\ref{apA}, we provide a quantitative analysis of the payoffs that shows that in the absence of defectors, the lowest contribution cooperators behave as defectors, while a higher contribution cooperator behaves like a cooperator with a different contribution value.

\subsubsection*{Coexistence and parasitism}
\label{paras}

\begin{figure}[t]
    \centering
    \includegraphics[width=\linewidth]{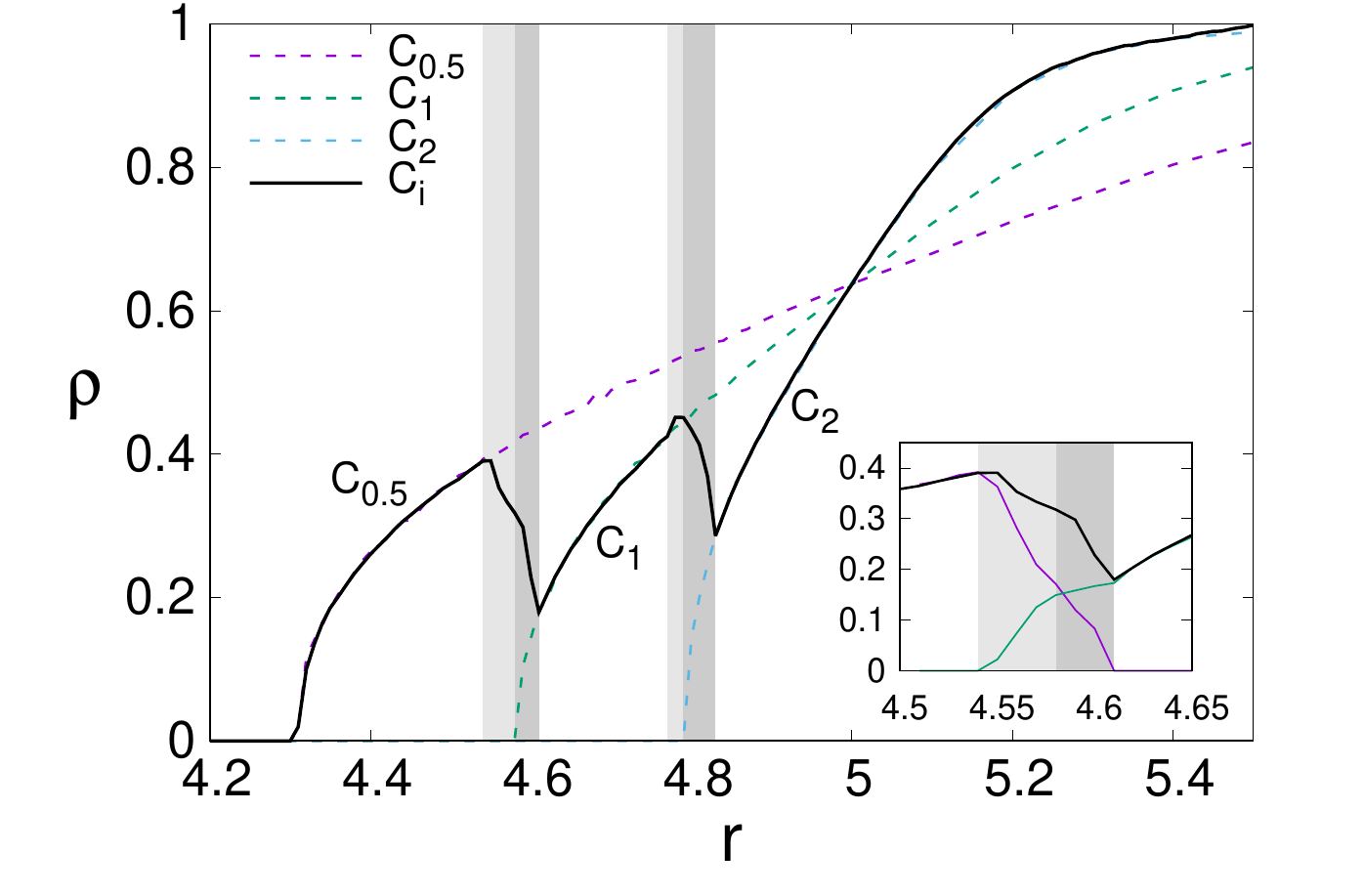}
    \caption{ Equilibrium density of cooperation as a function of $r$. The solid line corresponds to the heterogeneous case, where cooperators initially contribute with values of $0.5$, $1$, and $2$, uniformly distributed. The dashed lines represent the homogeneous cases for each individual contribution value. We observe that there are dominance regions for all possible $C$s. In the light-grey rectangles, we identify the parasitism regions, where  higher contribution cooperators survive at the expense of lower contribution ones, where they cannot survive on their own (see inset for the case between $C_1$ and $C_{0.5}$). 
    Specifically, the higher contribution $C$s can exploit any lower contribution cooperators, but for different values of $r$. However, in the dark-grey rectangles, coexistence regions appear where $C_1$ coexist with $C_{0.5}$, and $C_2$ can coexist with both. 
    }
    \label{coex2}
\end{figure}

Next, we will study a mix of different contribution cooperators in the presence of defectors.
In principle, different cooperators could coexist if $r$ is higher than both their homogeneous $r_c$ values. However, due to the behavior discussed previously, different $C$s can jeopardize one another.
The equilibrium density of cooperators as a function of $r$ is presented in Fig.~\ref{coex2} for both homogeneous cases (with $c=0.5$, $c=1.0$ and $c=2$, dashed curves) and the heterogeneous case (solid curve), where cooperators contribute the aforementioned values with equal probability in the initial configuration. 
First, when $r$ is only slightly above the critical value ($r_{c_o} \approx 4.3$) of the lowest contributing cooperator, $C_{0.5}$, all other contribution $C$s become extinct.
As $r$ approaches, but still remains below, the critical value ($r_c\approx 4.58 $) of $C_1$, a coexistence phase emerges for $C_{0.5}$ and $C_1$, despite the fact that the latter cannot survive alone. This is due to a parasitic behavior between them, which occurs in the first highlighted light grey region.
As we further increase $r$, at some point it will become higher than both homogeneous critical values of $C_{0.5}$ and $C_1$, characterizing the beginning of the coexistence phase between them. At some point for a high enough value of $r$, $C_1$ cooperators will be favored and coexist with defectors alone. 
 For $r$ close to, but still below the critical value ($r_c\approx 4.8$) of $C_2$ cooperators, a similar parasitism mechanism occurs and now $C_2$ cooperators take advantage of $C_{0.5}$ or $C_1$ cooperators (in different regions). 
Above all of the homogeneous $r_c$ values, we observe another coexistence region between $C_2$ cooperators and $C_{0.5}$ or $C_1$ (right dark grey rectangle). 

\begin{figure*}[t!]
\begin{center}
\includegraphics[width=.4\columnwidth]{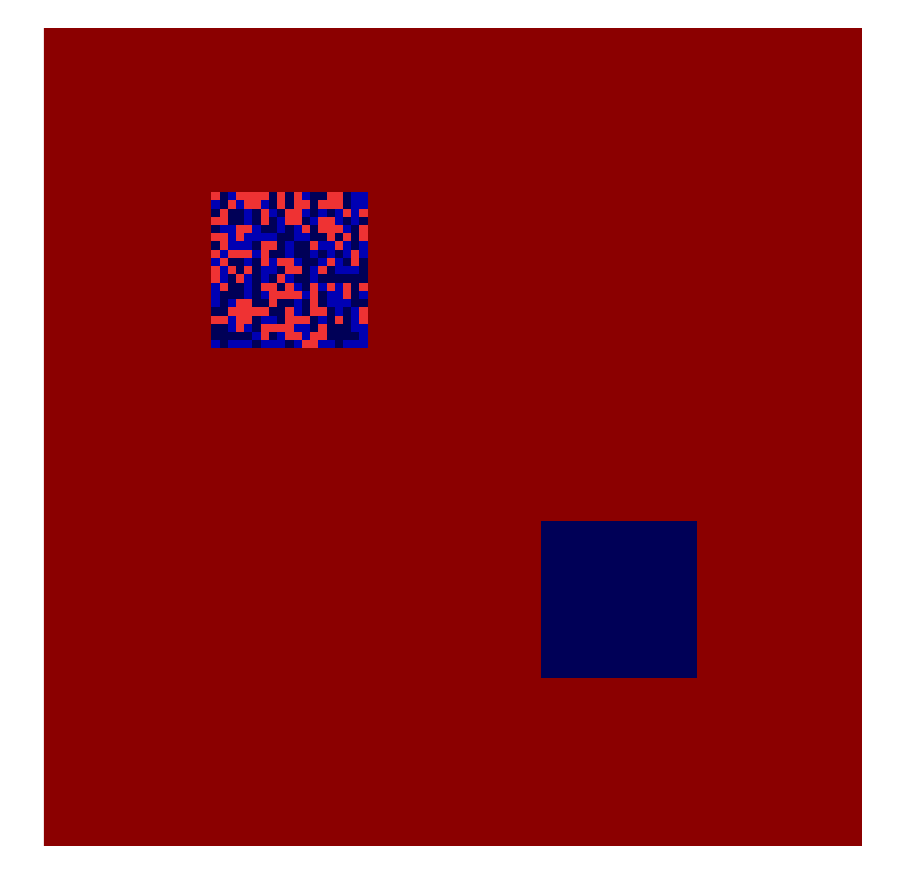}
\includegraphics[width=.4\columnwidth]{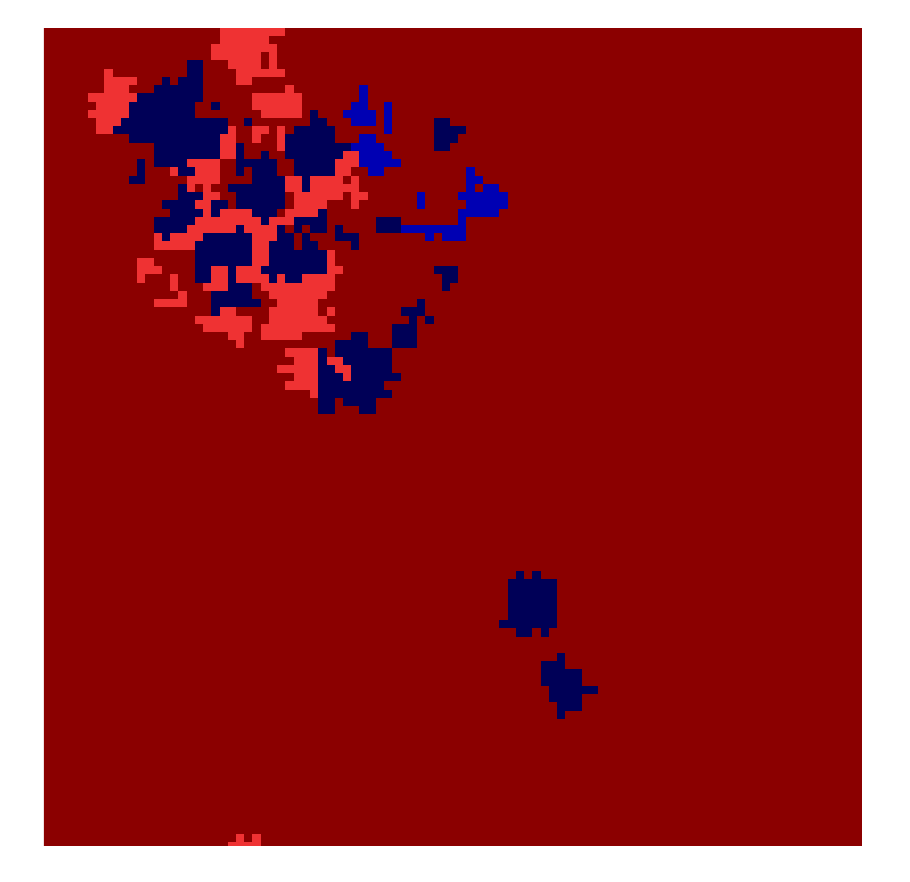}
\includegraphics[width=.4\columnwidth]{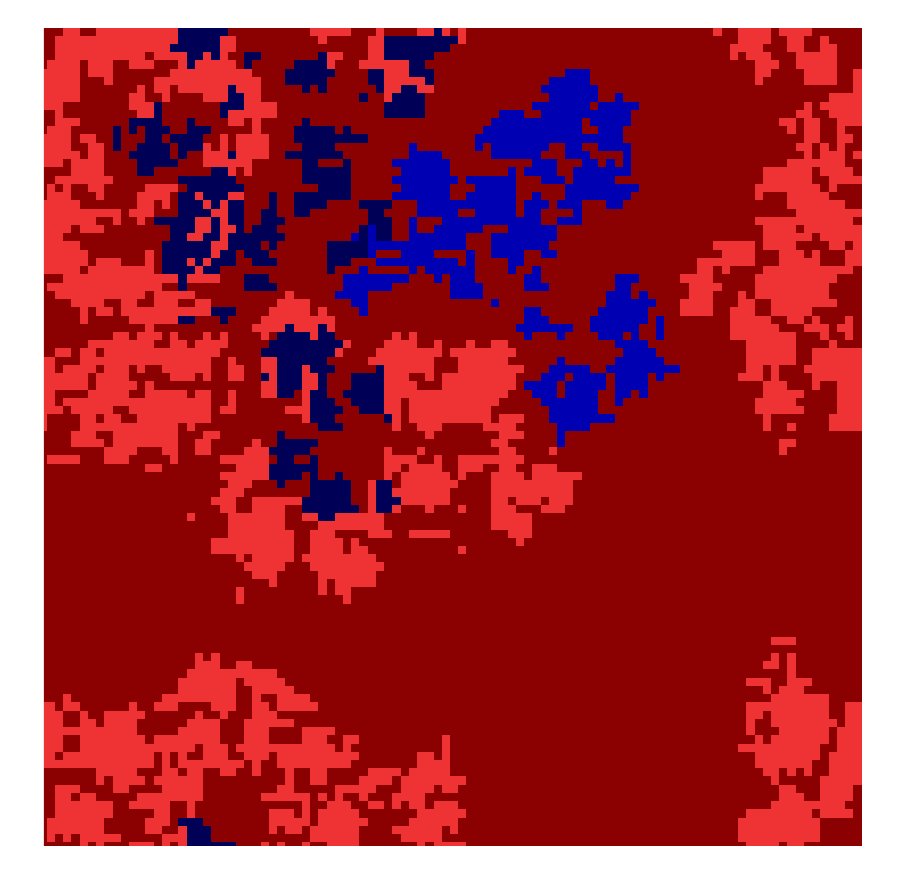}
\includegraphics[width=.4\columnwidth]{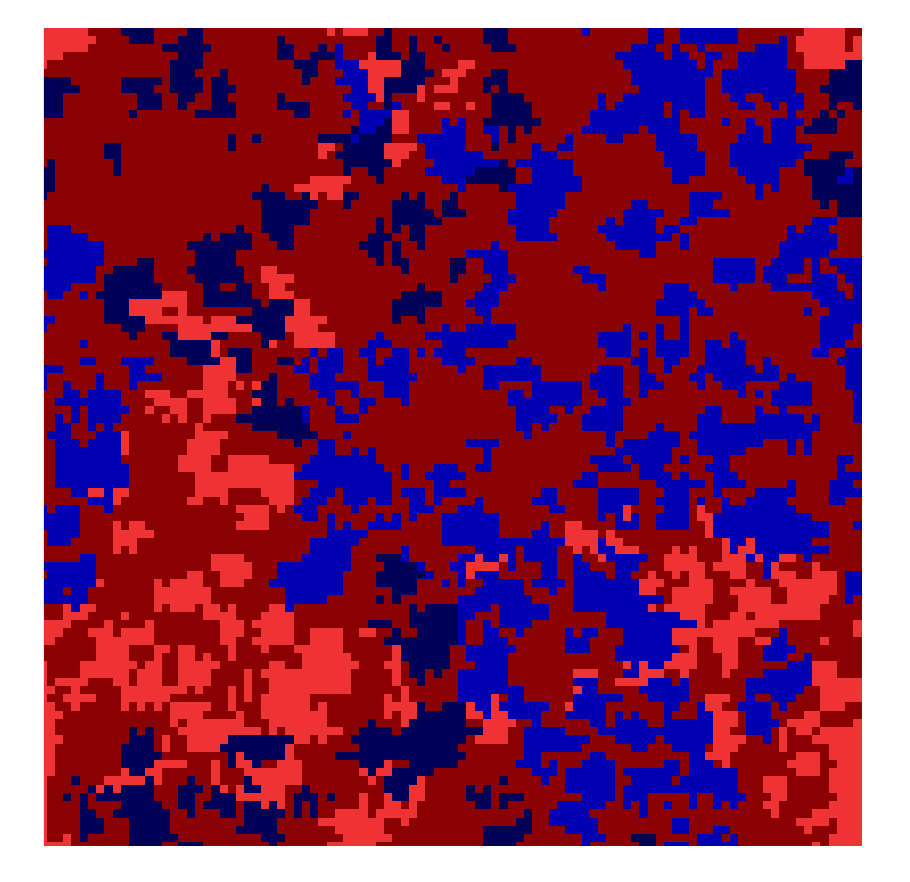}
\includegraphics[width=.4\columnwidth]{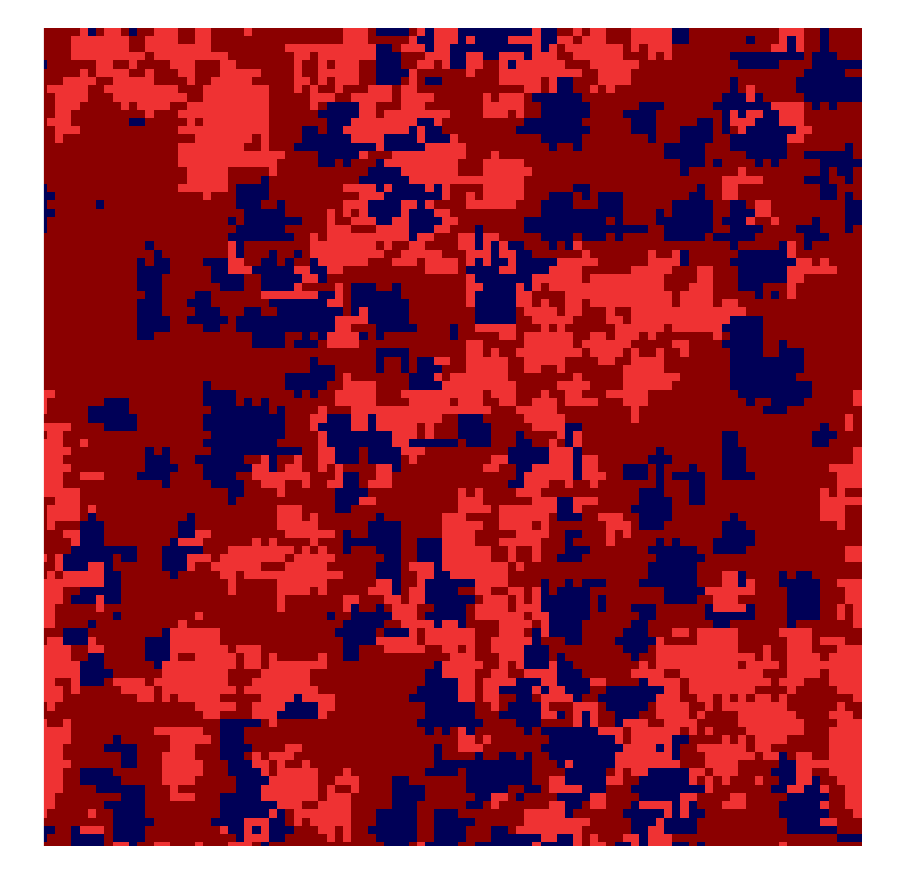}

\includegraphics[width=.4\columnwidth]{snapshot_MCS1-eps-converted-to.pdf}
\includegraphics[width=.4\columnwidth]{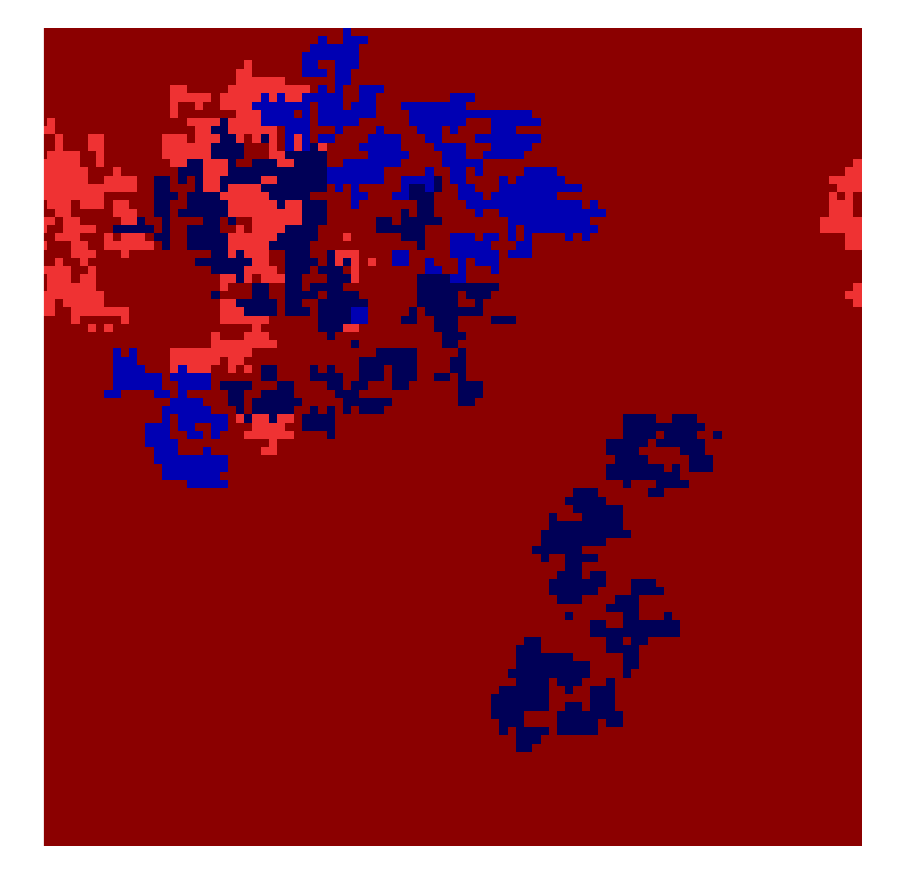}
\includegraphics[width=.4\columnwidth]{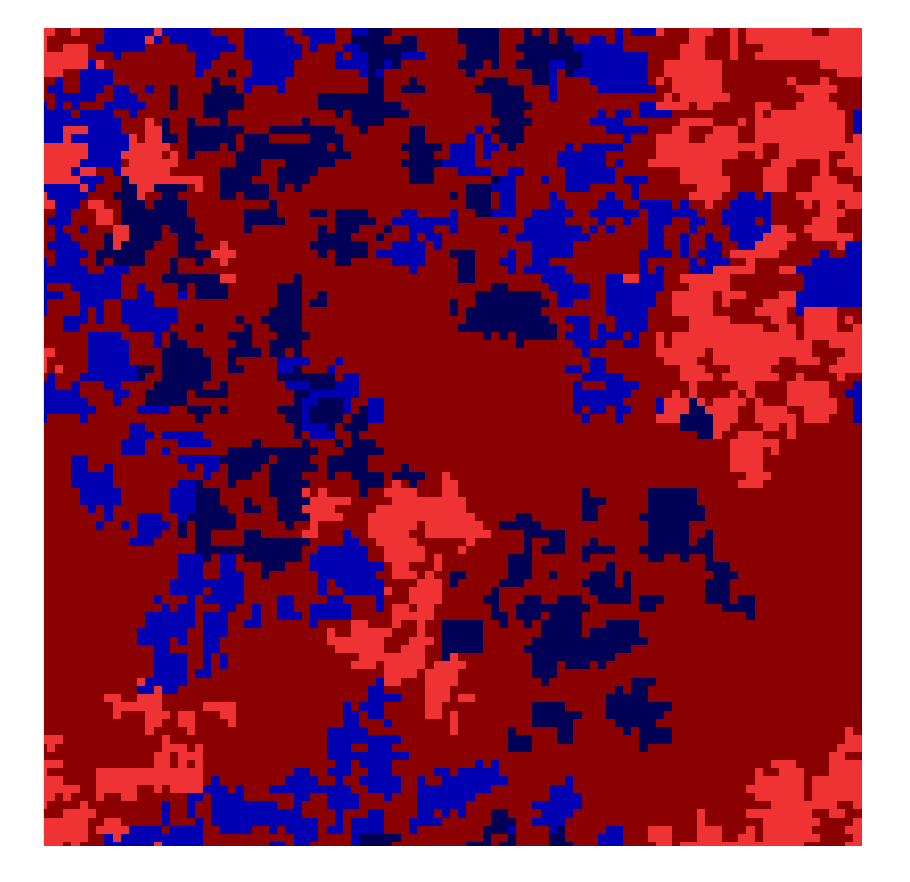}
\includegraphics[width=.4\columnwidth]{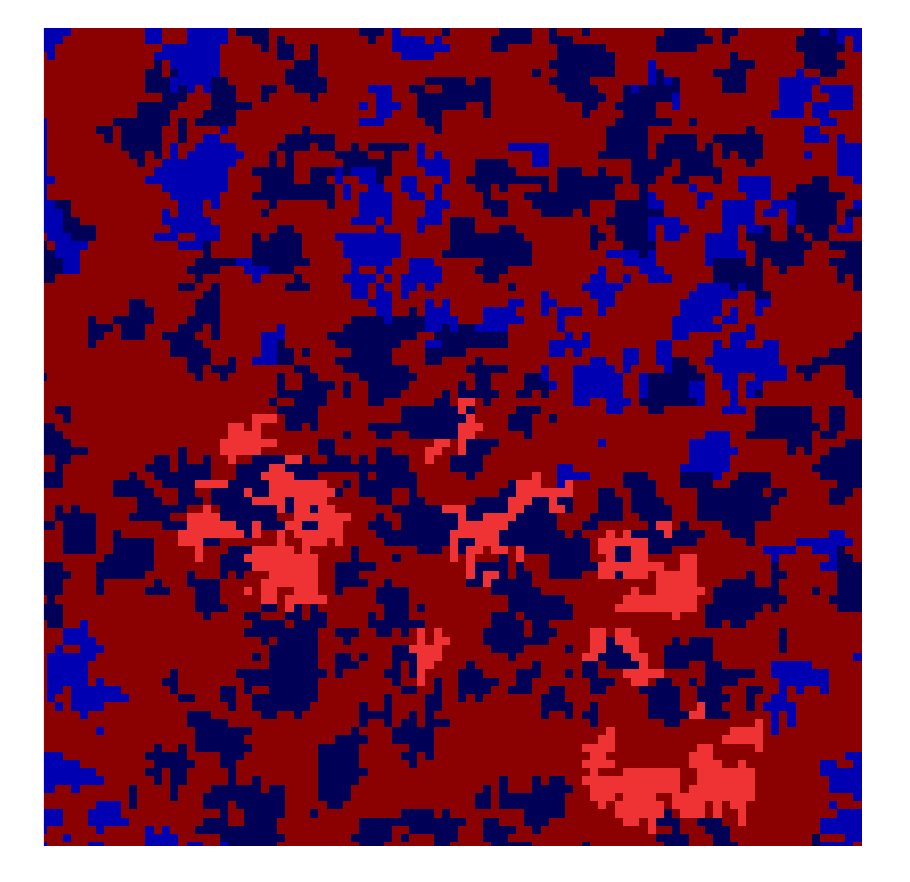}
\includegraphics[width=.4\columnwidth]{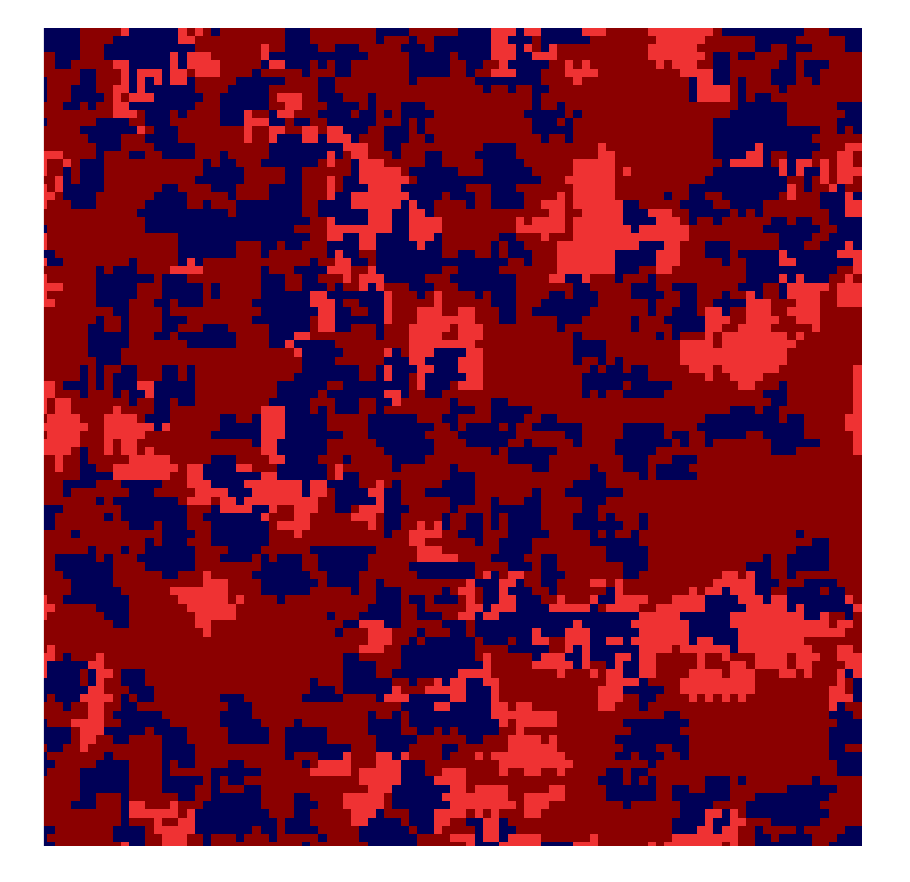}

    \caption{Time evolution of a cooperator clusters with $c=0.5$ (light red), $c=1.0$ (light blue) and $c=2$ (dark blue) in the sea of defectors (red) on the square lattice for simulation times $t = 1, 200, 500, 1000$ and $10000$ (MCS). The top row illustrates the parasitism region ($r=4.77$),  where high contribution cooperators ($C_2$) can only survive if they are in close proximity to low contribution $C_{0.5}$.  The bottom row shows the coexistence region ($r=4.81$), where the highest contribution cooperators, $C_{2}$ survive both in contact with the $C_{0.5}$ and on their own. 
    }
    \label{corte2}
    \end{center}
\end{figure*}

From the inset of Fig.~\ref{coex2}, it is clear that the higher contribution $C$s benefit from  the lower contribution ones.  Besides being able to survive in regions where they would be extinct when only in the presence of defectors,  we observe that their presence decreases the density of  $C_{0.5}$ when compared to their homogeneous equilibrium density. Overall, this lowers the total density of cooperators in this region.
In addition, Fig.~\ref{coex2} indicates that the critical $r$ value required to maintain cooperation in a heterogeneous scenario is determined by the lowest homogeneous $r_{c_i}$ value among all cooperators, since when $r$ is low all cooperators with a homogeneous $r_c>r$ cannot survive. 
This finding is  consistent with a similar model that used uniformly distributed contributions taken from $U(1-\sigma,1+\sigma)$~\cite{huang2015effect}, where $U(a,b)$ represents a continuous uniform distribution in the interval $[a,b]$. In that study, increasing $\sigma$, which eventually makes $U$ encompass our optimal contribution value ($c_0\approx 0.4$), reduced the critical $r$ required to sustain cooperation until it eventually plateaued for a sufficiently high $\sigma$ value.

Fig.~\ref{corte2} shows snapshots of the time evolution for the parasitism and coexistence between $C_2$ and $C_{0.5}$ cooperators. In the first row ($r=4.77$), we show the parasitic behavior where $C_2$ cooperators close to the lower contribution $C_{0.5}$ survive while an isolated cluster of $C_2$ is driven to extinction. 
The second row shows the region where $r$ is greater than all homogeneous  critical $r_c$ values($r\gtrsim 4.8$). While $C_2$ cooperators survive on their own because there are no higher contribution $C$s to hinder their growth, they will harm all other lower contribution $C$s. 
To see which other cooperators survive with them, we can refer to the case with fewer strategies for the same $r$ region.  With only $D$, $C_1$ and $C_2$ in the population, the $r$ values in question are high enough to favor $C_2$ with the exception of $r=4.8$ (data not shown). For this value, $C_1$ cooperators can coexist with $C_2$.
However, in the case $D$, $C_{0.5}$ and $C_2$ (data not shown), the $r$ values are still in the coexistence region between $C_{0.5}$ and $C_2$.
Therefore, $C_2$ can coexist with both strategies for $r=4.8$ but only with $C_{0.5}$ for higher values of $r$.

\subsection{Homogeneous vs heterogeneous cases}

By comparing the homogeneous cases with the heterogeneous one, represented by the dashed and solid lines in Fig.~\ref{coex2}, respectively, we observe that heterogeneity can favor cooperation, leading to a higher equilibrium density for $r<5$, when compared to the homogeneous case with $c=2$. 
This is due to the high density of the lower contribution cooperators that compensate the low density of high contribution cooperators when alone. On the other hand, when compared with the homogeneous case with $c=0.5$, the presence of high contribution cooperators prevents the expansion of lower contribution ones, resulting in a worse scenario for cooperation.

\begin{figure}[t]
    \centering
   \includegraphics[width=\linewidth]{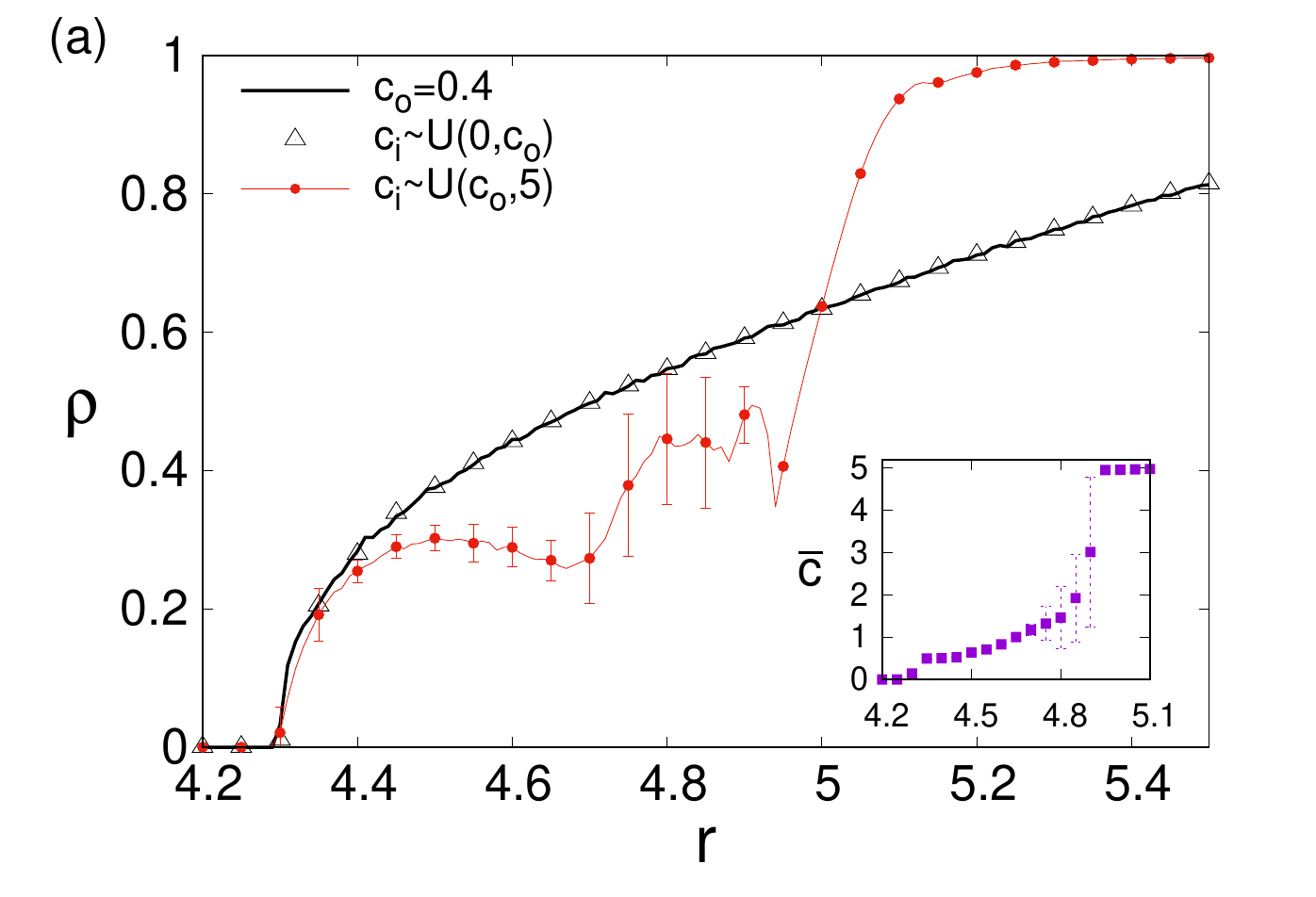}
   \centering
   \includegraphics[width=\linewidth]{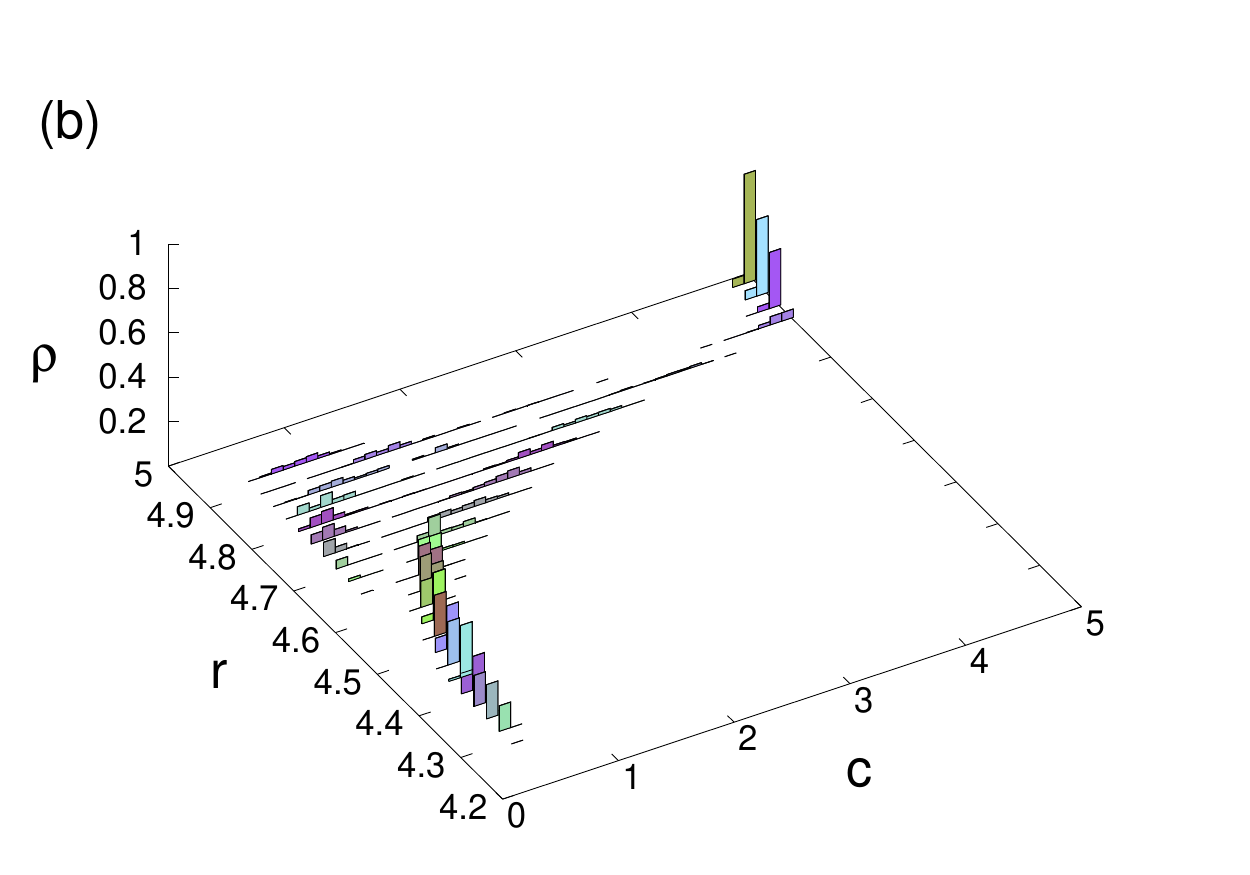}
    \caption{Equilibrium density of cooperators as a function of $r$ for different initial contribution distributions. 
    We observe that for contribution ranges that encompass the optimal homogeneous case $c_0$ and lower values (triangles), the equilibrium density is exactly the same as that of the homogeneous optimal case (solid black line). However, for distributions with contributions higher than the optimal (red circles), cooperation is inhibited for intermediate $r$ values ($4.4\lessapprox r <5$), when compared to the optimal. Moreover, in this case, there is great variability among samples, as indicated by the large standard deviations. 
     Here, $U(a,b)$ represents a continuous uniform distribution in the range $[a, b]$.
    The inset shows the mean contribution vale and the standard deviation from it between samples in the equilibrium for the $U(c_0,5)$.
    We also show in (b) all contributions that survived for each sample for the same distribution. While for low $r$ values only one contribution survive, for high enough values two or even three contributions start to survive together.
    }
    
    \label{unif}
\end{figure}

In summary, determining whether heterogeneity is beneficial for cooperation depends on the reference point. To obtain a more definitive answer, it is useful to compare the heterogeneous case to the homogeneous case at the optimal value. Fig.~\ref{unif} illustrates this scenario for various initial distributions. 
Near the critical point ($r_{c}\approx 4.3$), high-contribution cooperators do not hinder the optimal $C$s  since they are unable to survive at such low $r$ values. However, for distributions that do not encompass optimal cooperators, the critical $r$ will be higher, as previously discussed. 
As $r$ increases, higher contribution $C$s begin to survive and can jeopardize cooperation, as shown in the figure.

As demonstrated in Fig.~\ref{coex2}, various coexistence, parasitism, and dominance regions exist for different $C$s. As a result, for continuous distributions of the initial contributions, we expect a superposition of all possible scenarios, making predictions for each sample challenging, as indicated by the large standard deviations of the cooperator's equilibrium density in Fig.~\ref{unif}. 
In the inset of Fig.~\ref{unif}, we present the mean equilibrium contribution among samples and its standard deviation for the uniform $U(c_{o},5)$ case. Both of these quantities increase as $r$ increases. For low $r$ values, only one type of $C$ with a contribution close to the optimum survives. However, as $r$ increases, higher contribution $C$s become viable, leading to a decrease in the density of cooperation compared to the homogeneous optimal case.
For intermediate $r$ values, a range of contributions can survive, which also increases with $r$. Now, contributions closer to the optimal value survive for some samples, explaining the increase in the average density of cooperation. For high enough values of $r$, only the highest contribution is favored and therefore survives alone, matching the homogeneous $c=5$ curve.
In Fig.~\ref{unif} (b), we also display the contributions that survive for each sample. Interestingly, we observe four distinct regions that describe the individual contributions: for $r<4.7$, only one contribution survives with defectors; for $4.7<r<4.8$, two contributions can coexist; for $4.8<r<4.9$, three contributions can coexist; for $r>4.9$, either two contributions coexist or one survives alone.
Despite this, all possible scenarios are worse than the homogeneous one at the optimal contribution value for cooperation below $r=5$. This behavior is consistent for other distributions, including Gaussian and the uniform $U(0,5)$ (data not shown). 
Therefore, in terms of equilibrium density and critical $r$ value, heterogeneity can only harm cooperation compared to the optimal homogeneous case. 

It is important to note that the results presented above are specific to the square lattice, but they can be expected for all lattices with a null clustering coefficient~\cite{perc2013evolutionary}, where optimal noise values exist. 
For lattices with a non-null clustering coefficient, the deterministic scenario is known to be the best for cooperation in terms of the critical $r$ value.
Therefore, the best scenario for cooperation is expected to be achieved with a sufficiently high contribution, which corresponds to  low noise and, consequently, to the smallest possible $r_c$ value. Above this value, the dynamics will not change as the deterministic scenario has already been established. As discussed previously, Fig.~\ref{diagrama} (b) confirms this behavior for the triangular lattice.
For the heterogeneous case, high contribution cooperators will generally be preferable for cooperation since the deterministic case is the best. It is important to remember that higher contribution $C$s have an advantage over lower contribution $C$s.
Now, the higher contribution $C$s that survive the initial steps and manage to cluster will always survive alone.
Therefore, compared to a homogeneous case, heterogeneity will benefit cooperation (by reducing $r_c$) if it allows for higher contributions.
However, compared to the optimal homogeneous case, heterogeneity will not matter since all the lower contributions will become extinct, and the higher contributions will also be in the optimal scenario.

\section{Conclusion}\label{conclusion}

The concept of heterogeneity has become a crucial area of study in evolutionary game theory. Researchers have delved into the various benefits of heterogeneity in diverse situations, including those explored in the paper. While our work aimed to investigate the positive effects of heterogeneity, we also uncovered a potential downside. Unlike other papers, our study revealed how heterogeneity can impede cooperation. By examining both the positive and negative aspects of heterogeneity, our work contributes to a deeper understanding of the complexities of evolutionary game theory. The findings of our study can provide valuable insights into the optimization of cooperative behavior in various scenarios, helping to enhance the overall efficiency and effectiveness of social systems.

In summary, changing the contribution value of cooperators using the imitation update rule is essentially equivalent to scaling the noise. As there is an optimal noise value for sustaining cooperation, there exists an optimal homogeneous contribution for cooperators that minimizes the critical $r$ value.
We found that for regular lattices with a null clustering coefficient, heterogeneity can be detrimental to cooperators in the optimal scenario. 
However, for regular lattices with a non-null clustering coefficient, heterogeneity will not harm optimal cooperators. Therefore, we conclude that the homogeneous optimal scenario is the best for cooperation compared to the heterogeneous case  studied in this work.
This is due to the fact that higher contribution cooperators jeopardize smaller ones, and can even parasite them.

It is well established that the deterministic case is the best scenario for the classic PGG with group interactions for all types of regular 
lattices~\cite{perc2013evolutionary}. Therefore, we expect that all regular lattices will exhibit a behavior similar to that of the triangular lattice in our model for the PGG.
Moreover, it has been shown that the Focal PGG can be mapped to the Prisoner's Dilemma game for certain parametrizations~\cite{FLORES2022112744, hauert2003prisoner}. However, by these approaches, different contributions are related to different group sizes, which must be considered when studying heterogeneity's in the PD game.

\section*{CRediT authorship contribution statement}

L. S. Flores contributed with Software and all authors contributed equally in Conceptualization,
Formal analysis and Writing.
 \section*{Acknowledgments}
L.S.F. thanks the Brazilian funding agency CAPES (Coordenação de Aperfeiçoamento de Pessoal de Nível Superior) for the Ph.D. scholarship. We used ChatGPT (\url{chat.openai.com}) and Grammarly (\url{www.grammarly.com}) to improve the quality of the written text in English. The simulations were performed on the IF-UFRGS computing cluster infrastructure.

\appendix

\section{}
\label{apA}

Here, we show that cooperators with the lowest contributions behave as defectors in their absence.
We start from Eqs.~(\ref{eqC}) and (\ref{eqD}) for the classical FPGG, where all cooperators equally contribute $c$  and write the payoff difference for a given configuration
\begin{align}
\Pi_{C} -  \Pi_{D} &=  \frac{rc}{G} \left( N_{C}^{C} - N_{C}^{D}     \right) - c \\
    &= c \left( \frac{r}{G} \Delta N_{C} - 1 \right),  \label{eq1}
\end{align}
where $\Delta N_{C}$ is the difference in the number of cooperators between the two groups. 

Now, suppose that  instead of $C$ and $D$, we have only two types of interacting cooperators, $C_{c_i}$ and  $C_{c_j}$, where the subscripts denote their contribution value and $\Delta c = c_i-c_j >0$. In this case, the payoffs are
\begin{equation}
    \Pi_{C_{c_k}} = \frac{r}{G} \left(N_{c_i}^{k} c_i + N_{c_j}^{k} c_j \right) - c_k,
\end{equation}
where $N_{c_m}^{k}$ is the number of neighbors contributing $c_m$ in the group where $C_{c_k}$ is at the central site, for $k\in \{i,j\}$. 
The payoff difference for this situation is 
\begin{equation}
    \Pi_{C_{c_i}} -  \Pi_{C_{c_j}} = (c_i  \Delta N_{c_i}  + c_j  \Delta N_{c_j} ) \frac{r}{G} - \Delta c,
    \label{eq.dif}
\end{equation}
which generalizes Eq.~\ref{eq1} for two types of contributions. 

\begin{figure}[h]
    \centering
    \includegraphics[width=\linewidth]{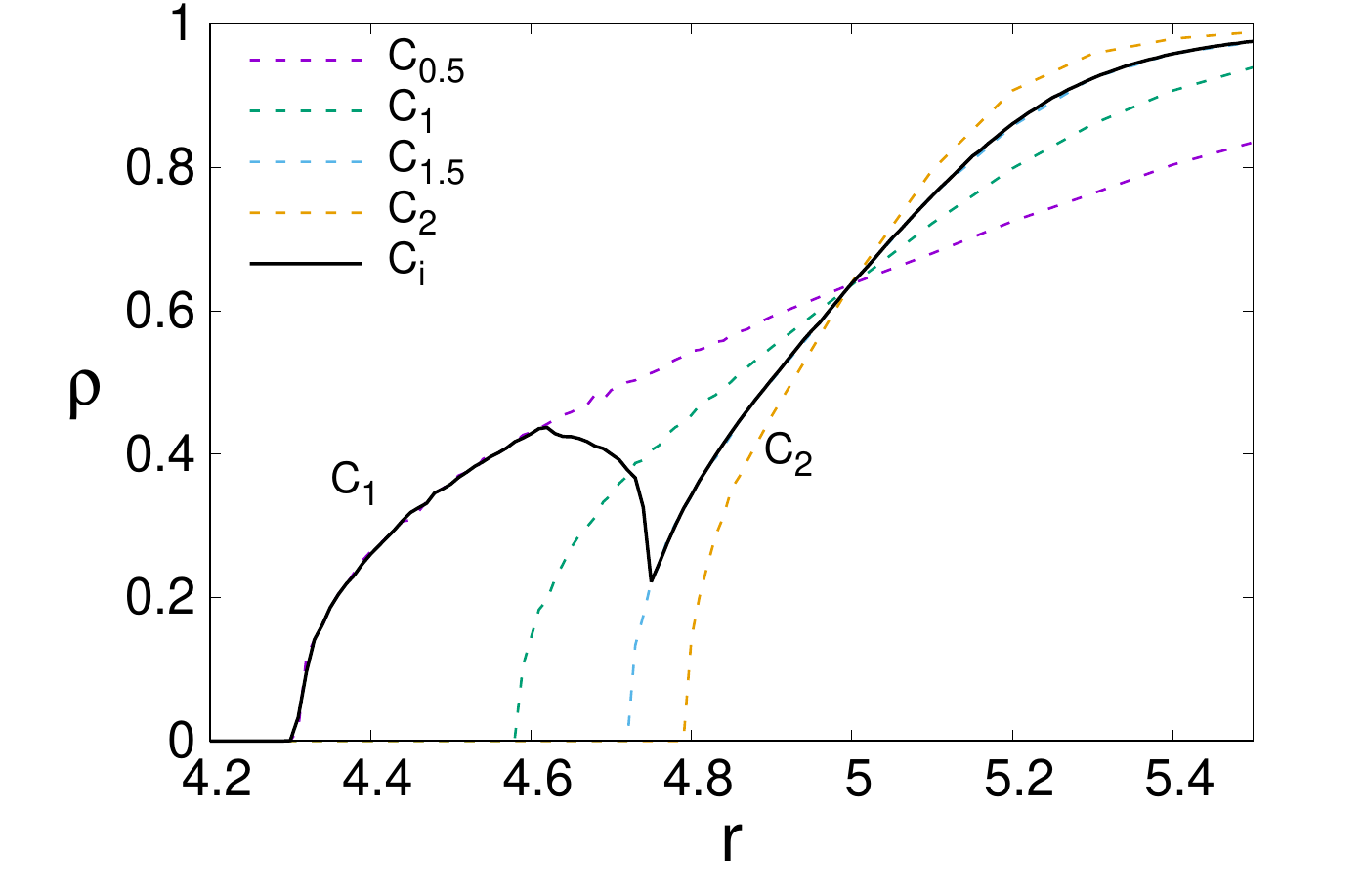}
    \caption{Equilibrium density of $C_{1}+C_{2}$ in function of $r$ for only cooperators in the population with contributions $0.5$, $1$ and $2$.  We observe in more details the situation explored in Fig.~\ref{corte}, showing all transitions between cooperators where the smaller contribution behaves as a defector and the other contributions as cooperators with different contributions than the original ones.  }
    \label{app}
\end{figure}

Considering the same spatial configuration as in the $C$ \textit{vs} $D$ case, 
 but with defectors replaced by  $C_{c_j}$, and  cooperators by $C_{c_i}$ 
we  have  $\Delta N_{c_i} = \Delta N_{C}$ by construction. 
To obtain $\Delta N_{c_j}$, we recall that the total number of players in a group is $G$ (including the focal player) and, therefore, 
\begin{equation}
    N_{c_i}^{k} + N_{c_j}^{k} = G. 
\end{equation}
 Then, by subtracting one of these equations from the other ($k\in\{i,j\}$), we  obtain 
\begin{equation}
      \Delta N_{c_i} = -  \Delta N_{c_j}.
\end{equation}
Therefore, 
 $ \Delta N_{C} =  \Delta N_{c_i} = -  \Delta N_{c_j}$ always holds.  Finally, with this result, we can rewrite Eq.~\ref{eq.dif} as
\begin{equation}
     \Pi_{C_{c_i}} -  \Pi_{C_{c_j}} = \Delta c \left( \frac{r}{G} \Delta N_{C} - 1 \right). \label{eq2}
\end{equation}
By comparing Eqs.~\ref{eq1} and \ref{eq2}, we see that the lower contribution $C_{c_j}$ behaves as a defector and the higher contribution $C_{c_i}$ behaves as a cooperator whose contribution is $\Delta c$.  
We illustrate this behavior in Fig.~\ref{app} for the case with only cooperators that contribute $c=0.5$, $c=1.0$ and $c=2.0$ in the population (in the absence of defectors).
We see that when $C_{1}$ cooperators coexist with $C_{0.5}$ we have $\Delta c = 0.5$ and therefore the density curve matches the homogeneous $C_{0.5}$ with defectors case. When $C_{2}$ cooperators coexist with $C_{0.5}$, $\Delta c = 1.5$ and therefore the density curve matches the homogeneous $C_{1.5}$ case.

\end{document}